\documentclass[preprint]{agujournal}
\draftfalse

\usepackage{bm}

\journalname{}

\begin{document}

%
%


\title{Maximizing simulated tropical cyclone intensity with action minimization}

%
%




\authors{David A. Plotkin\affil{1}, Robert J. Webber\affil{2}, Morgan E O'Neill\affil{3}, Jonathan Weare\affil{2}, Dorian S. Abbot\affil{1}}


\affiliation{1}{Department of the Geophysical Sciences, The University of Chicago, Chicago, Illinois, USA.}
\affiliation{2}{Courant Institute of Mathematical Sciences, New York University, New York, NY, USA.}
\affiliation{3}{Department of Earth System Science, Stanford University, Stanford, CA}




\correspondingauthor{David A. Plotkin}{dplotkin@intersystems.com}




\begin{keypoints}
\item Action minimization reduces the computational cost of intense cyclone simulation as compared to an ensemble study
\item Action minimization perturbs climate/weather models into producing realistic and more intense tropical cyclones
\item Time-dependent, asymmetric heating can contribute significantly to tropical cyclone intensification
\end{keypoints}

%
%


\begin{abstract}

Direct computer simulation of intense tropical cyclones (TCs) in weather models is limited by computational expense. Intense TCs are rare and have small-scale structures, making it difficult to produce large ensembles of storms at high resolution. Further, models often fail to capture the process of rapid intensification, which is a distinguishing feature of many intense TCs. Understanding rapid intensification is especially important in the context of global warming, which may increase the frequency of intense TCs. To better leverage computational resources for the study of rapid intensification, we introduce an action minimization algorithm applied to the WRF and WRFPLUS models. Action minimization nudges the model into forming more intense TCs than it otherwise would; it does so via the maximum likelihood path in a stochastic formulation of the model, thereby allowing targeted study of intensification mechanisms.

We apply action minimization to simulations of Hurricanes Danny (2015) and Fred (2009) at 6 km resolution to demonstrate that the algorithm consistently intensifies TCs via physically plausible pathways. We show an approximately ten-fold computational savings using action minimization to study the tail of the TC intensification distribution. Further, for Hurricanes Danny and Fred, action minimization produces perturbations that preferentially reduce low-level shear as compared to upper-level shear, at least above a threshold of approximately $4 \mathrm{\ m \ s^{-1}}$. We also demonstrate that asymmetric, time-dependent patterns of heating can cause significant TC intensification beyond symmetric, azimuthally-averaged heating and find a regime of non-linear response to asymmetric heating that has not been extensively studied in previous work.

\end{abstract}

%
%

%


%
%
%
%

\section{Introduction}
\label{sec:intro}


Despite extensive study, a consensus has yet to emerge about the mechanisms responsible for tropical cyclone (TC) intensification \citep{Mo2014}. The study of intensification remains challenging, owing to the number of scales and processes involved, e.g. mesoscale vortices, convection, turbulence, and microphysics \citep{Ro2013, Ka2015}, as well as the expense and danger of obtaining in situ observations. Because category three, four, and five TCs cause 85$\%$ of all TC damage in the United States despite comprising only $24\%$ of storms \citep{Pi2008b}, the study of these relatively less frequent storms is of particular interest. Further, many studies suggest that global warming will increase the frequency of the most intense TCs \citep{Kn2010, Em2013}. Understanding the intensification mechanisms of the strongest TCs is thus both especially important and especially computationally expensive, since it requires TC simulations that are both highly resolved enough to capture critical processes and numerous enough to provide sufficient sample size \citep{Mu2012, Ro2015}.

\citet{Ka2003} showed that $83\%$ of North Atlantic category four and five TCs undergo rapid intensification, defined as an increase of at least $15$ m s$^{-1}$ in the maximum sustained surface wind speed over 24 hours (as compared to $31\%$ of all TCs). The study of intense TCs therefore hinges upon developing a deeper understanding of the rapid intensification process. While rapid intensification may not be a distinct process but rather a tail event of the distribution of intensification rates \citep{Ko2015}, the computational difficulty of sampling this tail still presents a challenge to TC simulation.

Investigating the tail of the intensification distribution by direct computer simulation faces two principle difficulties: first, weather models require very high resolution to conclusively differentiate among proposed rapid intensification mechanisms \citep{Ge2010}. Second, very large ensembles are necessary to sample the most extreme events in the tail of the TC intensity distribution. In this study, we propose and implement a new methodology to study rapid intensification that circumvents these issues; it does so by targeting computational power at the problem of rapid intensification more specifically than can be done with direct simulation.

To better leverage computational resources for the problem of rapid intensification and thereby avoid the need for prohibitively many model runs, we apply an action minimization algorithm to the study of TC rapid intensification. The advantage of action minimization for the study of rapid intensification is that it allows for the targeted study of storms that experience the lowest pressure minima and the highest intensification rates. By targeting these storms, we substantially reduce the computational cost of accessing and studying the tail of the intensification distribution. The algorithm adds a series of perturbations to a model trajectory over time, nudging the model toward states with some characteristic of interest (in this case, an intense TC). Each perturbation is indistinguishable from noise and consists of an adjustment to each value in several two- or three-dimensional physical fields: zonal and meridional wind, temperature, surface pressure, water vapor mixing ratio, and geopotential.

We obtain the set of perturbations over time by solving an optimization problem - namely, minimization of the action functional. The action functional imposes costs both on perturbation size and on TC intensity (measured here by minimum central surface pressure) in the final state. We solve this optimization problem by leveraging the four-dimensional variational analysis (4D-Var) capability of the Weather Research and Forecasting (WRF) model \citep{Hu2009}. The optimization we solve seeks perturbations that are small and realistic, thus providing a physically plausible rapid intensification trajectory. By using a background covariance matrix generated by forward model runs, we ensure physically relevant perturbations. In fact, given an initial condition, action minimization yields perturbations that corresponding to the likeliest trajectory from that initial condition to an intense TC under a stochastically perturbed model. In this study, we take as initial conditions already formed TCs of intermediate intensity; however, in principle it should be possible to apply action minimization to a quiescent atmosphere to study genesis as well as intensification.

While our methodology uses software tools originally built for the purpose of data assimilation with 4D-Var, we solve a fundamentally different problem. The purpose of 4D-Var is to improve the agreement between a model trajectory and previously observed data (e.g., substates of the full state of the weather at several points in time), thereby improving estimates of weather states that actually occurred \citep{Zu1996}. In contrast, action minimization seeks realistic model trajectories to some hypothetical state (or set of states) of interest, thereby improving our understanding of how a model transitions to such states. Instead of trying to fit the model to observations, the goal is to find trajectories that lead to as great a pressure depression at the final time as is physically realistic. That said, the optimization problem we solve in this study follows closely the formulation used in weakly-constrained 4D-Var (where perturbations are allowed to the initial condition and to subsequent states, as opposed to strongly-constrained 4D-Var where perturbations are added only at the initial time \citep{Da2000, Ub2000}). We note that equation~\ref{action functional old} in section~\ref{sec:act} differs from equation 6 of \citet{Tr2006} only in the choice of the final cost function $\Phi$. In fact, when final cost $\Phi$ is always non-negative, our scheme is exactly weakly constrained 4D-Var with a single synthetic observation of $\sqrt{\Phi}$ taken at the terminal time and with value equal to zero. This interpretation fails if $\Phi$, can take negative values, as would occur, for example, if we were to attempt to maximize wind speed (or any other non-negative quantity). In this case we would choose $\Phi$ to be negative wind speed.  We give a natural (and general) statistical interpretation of action minimization below in section~\ref{sec:act}. Nonetheless, the close mathematical relationship between action minimization and 4D-Var is key for our work, as it allows us to leverage the tangent linear and adjoint capabilities of the WRF model 4D-Var system, which are discussed in more detail in section~\ref{wrfplus}. In this study, we investigate the processes by which simulated storms undergo rapid intensification to form intense TCs; we do not aim to provide information about the history of specific storms.
 
Our work builds on that of \citet{Ho2006b} and \citet{Ho2006c}, which first applied algorithms based on 4D-Var to optimal TC perturbation. These studies were both done in a geoengineering context; \citet{Ho2006b} sought perturbations to reduce cyclone damage based on wind speeds and property values, while \citet{Ho2006c} sought perturbations for deflecting cyclone tracks away from population centers. These studies were done at 20 km resolution with a diagonal background covariance matrix, allowed perturbations only to the initial condition, relied on TC bogusing (artificially embedding an axisymmetric vortex into a regional weather model), and in the case of \citet{Ho2006c} constrained the model trajectory to be near a specific target state. While related, our work differs in that it seeks perturbations that provide physical insight into the intensification process. We achieve this by maximizing storm intensity at 6 km resolution, using a background covariance matrix that enforces physically realistic perturbations, allowing perturbations throughout the model trajectory, and using storms that arise from our choice of initial and boundary conditions.

We note as well that there is extensive literature that uses data assimilation techniques, especially the ensemble Kalman filter (EnKF), for the study of tropical cyclones \citep{Ka1960, Ev2003}. For example, \citet{Ch2007} used EnKF to improve estimates of vortex position while \citet{Po2014} assimilated dropsonde measurements to improve predictability of Hurricane Karl. Unlike 4D-Var, the EnKF does not require either the computational expense or the linearization associated with running an adjoint model; it therefore may appear advantageous to build an algorithm to sample extreme TC events based upon the EnKF instead of on 4D-Var. However, the Gaussian approximation of the forecast distribution upon which the EnKF relies is known to be violated for observations lying far into the tail of the forecast, leading to substantial assimilation errors in that case (see \citet{Mi1994, Mi1999} and \citet{Va2013}). For exactly the same reasons, one should not reply upon a Gaussian approximation of the  forecast distribution in a method that attempts to generate events far into the tail of that distribution.  We note however, that (like 4D-Var) action minimization does require that we introduce a statistical model for small perturbations (model error) added to the deterministic dynamics, and that we have chosen to model those perturbations as Gaussian random variables.  Adding a small Gaussian perturbation to a random variable (in this case the forecast state) has the effect of slightly smoothing the distribution of that random variable but leaving large scale features of the distribution (e.g., multimodality, skewness, etc.) intact, at least when the perturbations are small. For this reason, though the relative value of a tail probability may depend sensitively on the details of the perturbations, we expect that the maximum likelihood path found by action minimization (see section~\ref{sec:act}) will be more robust to those details.

The purpose of this study is to demonstrate the feasibility of using action minimization to study TC rapid intensification and to investigate the action-minimizing perturbations for information about the mechanisms of intensification. We apply action minimization to simulations of two category three hurricanes, Danny (2015) and Fred (2009), that were near the threshold for rapid intensification. We argue that action minimization causes both storms to undergo rapid intensification via physically realistic pathways. We also compare the computational cost of action minimization to that of studying these intensification pathways with ensemble simulation, showing an estimated ten-fold reduction in computational cost resulting from action minimization. We conclude that the extreme events at the tail of the TC intensification distribution can be usefully interrogated using action minimization.

For hurricanes Danny and Fred, the optimal perturbations show that reduction of low-level shear (the difference between horizontal winds at 900- and 500-hPa levels), at least above a threshold of approximately $5 \mathrm{\ m \ s^{-1}}$, has a much stronger effect on intensification compared to reduction of high-level shear (the difference between horizontal winds at 850- and 250-hPa levels). This is consistent with the findings of \citet{Fi2016}, who find that deep shear is particularly inhibitive to intensification since it tilts storms farther into the downshear-left quadrant (that is, to the left of the shear vector). Our results support the findings of \citet{Wo2004} and \citet{No2012} in that they suggest that there is an upper threshold above which shear is highly detrimental to intensification but not necessarily a lower threshold below which reducing shear is also detrimental to intensification. Further, we show that asymmetric, time-dependent patterns of heating can cause significant TC intensification beyond symmetric, azimuthally-averaged heating. In doing so, we find a regime of non-linear response to asymmetric heating that has not been extensively studied in previous work. The patterns of heating produced by action minimization are consistent with recent observational and modeling studies that posit the importance of relatively low-level latent heating for intensification \citep{Pa2013, Za2014b, Ta2015}. The action-minimizing heating patterns also support work that argues that TC response to asymmetric heating can be nonlinear and depends significantly on pre-existing vortex asymmetries \citep{Mo2005}.

\section{Methods} \label{sec:methods}

\subsection{Action minimization} \label{sec:act}

Algorithms to find the most likely path of a stochastic dynamical system between two states (or sets of states) have been used to great effect in computational chemistry for two decades \citep{Jo1998, We2002, We2010}.  In most of those applications the goal is to study the very infrequent transitions between two long lived states by identifying a trajectory of the process connecting the two states that is exponentially more likely than other potential trajectories in the small temperature limit as detailed by the theory of large deviations \citep{Fr1984}.  The method that we apply here is closely related to the method in \citet{We2004}, though here we also allow perturbations of the initial condition and we do not perturb the process at every model time step.

In addition to \citet{Ho2006b} and \citet{Ho2006c}, several previous studies have explicitly pointed to potential impact of rare event simulation and analysis tools on geophysical applications: a similar path finding approach was considered in the context of rogue ocean waves by \citet{De2018}. Methods designed to generate random samples of rare events were considered in the contexts of rare transitions of an ocean current model in \citet{We2009} and \citet{Va2013} and extreme heat waves in \citet{Ra2018}.

Given a model of some physical process, we would like to add small perturbations to true model trajectories in order to achieve a transition into some rare state. The goal of the action minimization algorithm is to minimize the magnitude of the perturbations added to the model, thus obtaining the most physically realistic path from the initial condition to a state of interest (in our case, any state with a particularly intense TC). Similar to action minimization, the 4D-Var cost function penalizes perturbations away from the model trajectory; however, our final cost draws the model trajectory toward final states with intense TCs, whereas the 4D-Var final cost penalizes the squared distance from observations of states along the model trajectory.

The following quantities are necessary to perform the minimization:

\begin{eqnarray*}
{\bf x_i} &:& \parbox{40em}{The state of the system at time index i. Includes horizontal velocity, \\temperature, pressure, water vapor mixing ratio, and geopotential\\ at every gridpoint.}\\[.25em]
F(\cdot) &:& \parbox{40em}{The function that integrates the model state to the next time index. \\In the absence of perturbations, this gives ${\bf x_i} = F({\bf x_{i-1}})$.   } \\[.25em]
{\bf \hat{\eta}_i} &:& \parbox{40em}{The perturbation at time index i.} \\[.25em]
\Phi({\bf x_i}) &:& \parbox{40em}{A function (order parameter) that measures the intensity of a TC at state ${\bf x_i}$.}\\[.25em]
N &:& \parbox{40em}{The number of optimization time steps.}\\[.25em]
B &:& \parbox{40em}{The background covariance matrix.}\\[.25em]
R_i, R_f &:& \parbox{40em}{Constants that balance running cost against final cost.}\\[.25em]
{\bf a} &:& \parbox{40em}{The initial state ${\bf x_0}$ is restrained to remain near the state ${\bf a}$.}
\end{eqnarray*}

Importantly, N is the number of perturbations added to a model trajectory; it is not necessarily the same as the number of model time steps. In principle, perturbations can be added every time step; however, we use one perturbation for every hour of model time whereas the time step for model dynamics is $30$ seconds. This is done in order to avoid the memory cost of storing perturbations at each time step. We demonstrate in section~\ref{sec:sens} that this choice of perturbation frequency does not qualitatively affect the results.

We seek to minimize the perturbations added to a true model trajectory in order to obtain a trajectory that results in a lower value of our order parameter $\Phi({\bf x_N})$ at the final state. We define the action functional $J$ as:

\begin{linenomath*}
\begin{eqnarray}
J &=& \sum_{i=0}^{N-1} \frac{1}{2 R_i^2} {\bf \hat{\eta}_i}^T B^{-1} {\bf \hat{\eta}_i} +\frac{1}{R_f^2} \Phi \left({\bf x_N}\right) \label{action functional old} \\
{\bf x_i} &=& F({\bf x_{i-1}}) + {\bf \hat{\eta}_i}, \ \ \ \ \ \mathrm{0<i<N} \nonumber \\
{\bf x_0} &=& {\bf a} + { \bf \hat{\eta}_0} \nonumber
\end{eqnarray}
\end{linenomath*}

$J$ is large both if the perturbations ${\bf \hat{\eta}_i}$ are large (high running cost) or if the order parameter $\Phi \left({\bf x_N}\right)$ is large (high final cost). The order parameter $\Phi$ we choose is the minimum surface pressure at the final state. By minimizing $J$, we seek a model trajectory with both small perturbations (indicating a physically relevant intensification process) and a final state that has a center of low minimum surface pressure. Given a domain over the tropical North Atlantic, such a final state should correspond to an intense TC. Constants $\{R_i\}_{i=0}^{N-1}$ and $R_f$ determine the relative importance of the running cost at each time step and of the final cost. We choose them heuristically so as to get the maximal intensification without causing the model to crash.

The covariance matrix $B$ is obtained using the National Meteorological Center (NMC) method from \citet{Pa1992}, where the covariance is thus the time averaged outer product of the differences between $12$- and $24$-hour forecasts produced with the WRF model over the course of several months (here chosen to correspond with the Atlantic hurricane season). As noted in \citet{Xi2009}, this introduces some approximation since the correlations obtained in this manner are accurate for climatological data but not necessarily for TC conditions.

To improve the conditioning of equation \ref{action functional old}, we substitute $\boldsymbol{\eta_i} = B^{-\frac{1}{2}}{\bf \hat{\eta}_i}$, which yields: 

\begin{linenomath*}
\begin{eqnarray}
J &=& \sum_{i=0}^{N-1} \frac{1}{2 R_i^2} \boldsymbol{\eta_i}^T \boldsymbol{\eta_i} +\frac{1}{R_f^2} \Phi \left({\bf x_N} \right)  \label{action functional} \\
{\bf x_i} &=& F({\bf x_{i-1}}) +  B^{\frac{1}{2} } \boldsymbol{\eta_i}, \ \ \ \ \ \mathrm{0<i<N} \ \nonumber \\
{\bf x_0} &=& {\bf a} + B^{\frac{1}{2}} \boldsymbol{\eta_0} \nonumber
\end{eqnarray}
\end{linenomath*}

Minimizing $J$ requires the derivatives of both the running cost and the final cost with respect to the set of perturbations $\boldsymbol{\eta_i}$:

\begin{linenomath*}
\begin{equation}
\frac{\partial J}{\partial \boldsymbol{\eta_i}} = \frac{1}{R_i^2} \boldsymbol{\eta_i} +\frac{1}{R_f^2} \left( \frac{\partial {\bf x_N}}{\partial \boldsymbol{\eta_i}} \right)^T  \frac{\partial \Phi ( {\bf x_N})}{\partial {\bf x_N}} \label{deriv one} \ \ \ .
\end{equation}
\end{linenomath*}

where $\boldsymbol{\eta} = \{\boldsymbol{\eta_i}\}_{i=0}^{N-1}$ is the set of all perturbations at all times. We apply the chain rule to obtain the derivative of the final cost:

\begin{linenomath*}
\begin{eqnarray}
 \frac{\partial {\bf x_N}}{\partial \boldsymbol{\eta_i}}  &=&  \frac{\partial {\bf x_N}}{\partial {\bf x_i}}   \frac{\partial {\bf x_i}}{\partial \boldsymbol{\eta_i}} \label{dpsi} \\
 \frac{\partial {\bf x_i}}{\partial \boldsymbol{\eta_i}} &=& B^{\frac{1}{2}} \nonumber \ \ \ .
\end{eqnarray}
\end{linenomath*}

We thus need  $\frac{\partial {\bf x_N}}{\partial {\bf x_i}}$, i.e., the derivative of the final state with respect to each previous physical state ${\bf x_i}$. We formulate a recursion that simplifies calculation of the necessary derivatives:

\begin{linenomath*}
\begin{eqnarray}
{\bf x_0} &=& {\bf a} + { \bf \hat{\eta}_0} \label{psis} \\
{\bf x_1} &=& {\bf \hat{\eta}_1} + F({\bf x_0})   \nonumber \\
{\bf x_2} &=& {\bf \hat{\eta}_2} + F({\bf x_1}) = {\bf \hat{\eta}_2} + F\left( {\bf \hat{\eta}_1} + F({\bf x_0}) \right) \nonumber \\
{\bf x_3} &=& {\bf \hat{\eta}_3} + F({\bf x_2}) = {\bf \hat{\eta}_3} + F\left[ {\bf \hat{\eta}_2} + F\left(  {\bf \hat{\eta}_1} + F({\bf x_0}) \right) \right]  \nonumber.   
\end{eqnarray}
\end{linenomath*}

Using equation~\ref{psis}, we differentiate each ${\bf x_j}$ with respect to each ${\bf x_i}$, which makes clear the general form of $\frac{\partial {\bf x_N}}{\partial {\bf x_i}}$:

\begin{linenomath*}
\begin{eqnarray}
\frac{\partial {\bf x_3}}{\partial {\bf x_2}} &=& F'({\bf x_2})    \\
\frac{\partial {\bf x_3}}{\partial {\bf x_1}} &=& \frac{\partial {\bf x_3}}{\partial {\bf x_2}} \frac{\partial {\bf x_2}}{\partial {\bf x_1}} = F'({\bf x_2}) F'({\bf x_1})  \nonumber \\
\Rightarrow \frac{\partial {\bf x_N}}{\partial {\bf x_i} } &=& F'({\bf x_{N-1}}) \dots F'({\bf x_{i+1}}) F'({\bf x_i}) \nonumber
\end{eqnarray}
\end{linenomath*}

$F'({\bf x})$ is the derivative of the flow; that is, $F'$ measures how a change in initial state ${\bf x}$ affects the state obtained by integrating ${\bf x}$ forward with $F({\bf x})$. We substitute into equation~\ref{dpsi} to obtain the derivative of the final state ${\bf x_N}$ with respect to the perturbations $\boldsymbol{\eta_i}$:

\begin{linenomath*}
\begin{eqnarray*}
\left( \frac{\partial {\bf x_N}}{\partial \boldsymbol{\eta_i}} \right)^T =B^{\frac{1}{2}} \left[ F'({\bf x_i}) \right]^T  \left[ F'({\bf x_{i+1}}) \right]^T  \dots  \left[ F'({\bf x_{N-1}}) \right]^T \label{deriv two}
\end{eqnarray*}
\end{linenomath*}

Substituting back into equation~\ref{deriv one} yields the full derivative of action functional $J$ with respect to perturbations $\boldsymbol{\eta_i}$:

\begin{linenomath*}
\begin{equation}
\frac{\partial J}{\partial \boldsymbol{\eta_i}} = \frac{1}{R_i^2} \boldsymbol{\eta_i} + \frac{1}{R_f^2} B^{\frac{1}{2}}\left[ F'({\bf x_i}) \right]^T  \left[ F'({\bf x_{i+1}}) \right]^T  \dots  \left[ F'({\bf x_{N-1}}) \right]^T \left( \frac{\partial \Phi \left( {\bf x_N}\right)}{\partial {\bf x_N}}  \right) \label{full deriv} \ \ \ .
\end{equation}
\end{linenomath*}

We use equation~\ref{full deriv} and a conjugate gradient method to find the action-minimizing perturbations $\boldsymbol{\eta_i^*}$ of equation~\ref{action functional}. The trajectory of physical states $\{{\bf x_i^*} \}_{i=0}^N$ that corresponds to this solution is given then given by:

\begin{linenomath*}
\begin{eqnarray}
{\bf x_i^*} = \Bigg\{
\begin{array}{lr}
{\bf a} + B^{1/2} \boldsymbol{\eta_0^* } & i = 0\\[.3em]
F({\bf {\bf x^*_{i-1}} }) + B^{1/2} \boldsymbol{\eta_i^*} & 0 < i < N
        \end{array}
\end{eqnarray}
\end{linenomath*}

That is, the initial state is given by adding the $t = 0$ perturbation to the physical initial condition. Subsequent states ${\bf x_i}^*$ are obtained by integrating the previous state ${\bf x^*_{i-1}}$ and adding the corresponding optimal perturbation $\boldsymbol{\eta^*_i}$.

The optimal trajectory of states $\{{\bf x_i}^*\}_{i=1}^{N}$ can be considered as a particular realization of a stochastic version of the original forward model. This implied stochastic model has Gaussian noise with covariance $BR_i^2$ added at each time index $i$. Such a model formulation is nearly equivalent to that used in 3d- and 4D-Var \citep{Hu2009}, including for the study of TCs \citep{Si2011}, and in other common applications including ensemble simulation \citep{Pe2006}. This implied model will be used to evaluate the efficacy of action minimization as compared to ensemble simulation in section~\ref{sec:justification}. The use of small stochastic perturbations in ensemble simulations serves to diversify the set of simulated trajectories and is often justified as representing subgrid effects and other model errors \citep{Oo2006, Le2008}. The trajectory produced by action minimization is therefore physical in the same sense that those produced by an ensemble simulation of the same stochastic model are physical. In fact, our action minimization results in the maximum likelihood trajectory within the stochastic model conditioned on trajectories having equal or lower terminal values of the order parameter, i.e. on trajectories ${\bf x_i}$ with $\Phi({\bf x_N}) \leq \Phi({\bf x_N^*})$.  To see this, note that the likelihood of any trajectory under the stochastic model is proportional to \[e^{-\sum \hat{\boldsymbol{\eta}}^T B^{-1} \hat{\boldsymbol{ \eta}} / 2 R_i^2}.\] If a trajectory other than ${\bf x_i^*}$ has both higher likelihood and equal or lower value of $\Phi({\bf x_N})$ then it would have a smaller action than ${\bf x_i^*}$.

\subsection{WRFPLUS}
\label{wrfplus}

Equation~\ref{full deriv} shows that the derivative of the model flow, $F'({\bf x_i})$, is necessary to compute the derivative of the action functional; this in turn requires a model with adjoint capabilities. Among open source models with this capability, the WRF Variational Data Assimilation (WRFDA) model \citep{Hu2009} is state-of-the-art. While WRFDA is normally used for four-dimensional variational data assimilation \citep{Ba2012}, the module WRFPLUS contains stand-alone tangent linear and adjoint versions of WRF as obtained by automatic differentiation and subsequent by-hand correction. We use the adjoint model within WRFPLUS to obtain derivatives of the model flow.

The adjoint model takes as input a perturbation vector $\boldsymbol{\eta}$ and state ${\bf x}$ and returns sensitivity $F'({\bf x})^T \cdot \boldsymbol{\eta}$, which is valid at the previous time index. Sensitivities are available for a subset of the physical variables, and the optimization is performed over perturbations with respect to these variables. Specifically, we optimize over perturbations to ${\bf u}$ (zonal wind), ${\bf v}$ (meridional wind), ${\bf p_{\mathrm{sfc}}}$ (surface pressure), ${\bf T}$ (temperature), ${\bf r}$ (water vapor mixing ratio), and ${\bf \phi}$ (geopotential, which describes the gravitational potential energy per unit mass at a given height).

\subsection{Experimental design}
\label{sec:exp}

To test the ability of the algorithm to force TCs within WRF to undergo rapid intensification, we chose two category three Atlantic TCs as test cases: 1) Hurricane Danny (2015), and 2) Hurricane Fred (2009). We use these storms as they were of similar intensity and stayed over open ocean, allowing for a simpler proof of concept. Both storms were also close to but did not quite meet the threshold of rapid intensification, making them attractive test cases. Simulations for both storms were done at 6 km resolution with a 36 hour time horizon chosen such that peak intensity is attained at the end of the simulation. The goal of our simulations was to cause the two two storms to intensify more than they would (ideally, close to maximum potential intensity discussed in section~\ref{MPI}) in the absence of action minimization.

\begin{figure}[h]
\centering
\includegraphics[width =.85\textwidth]{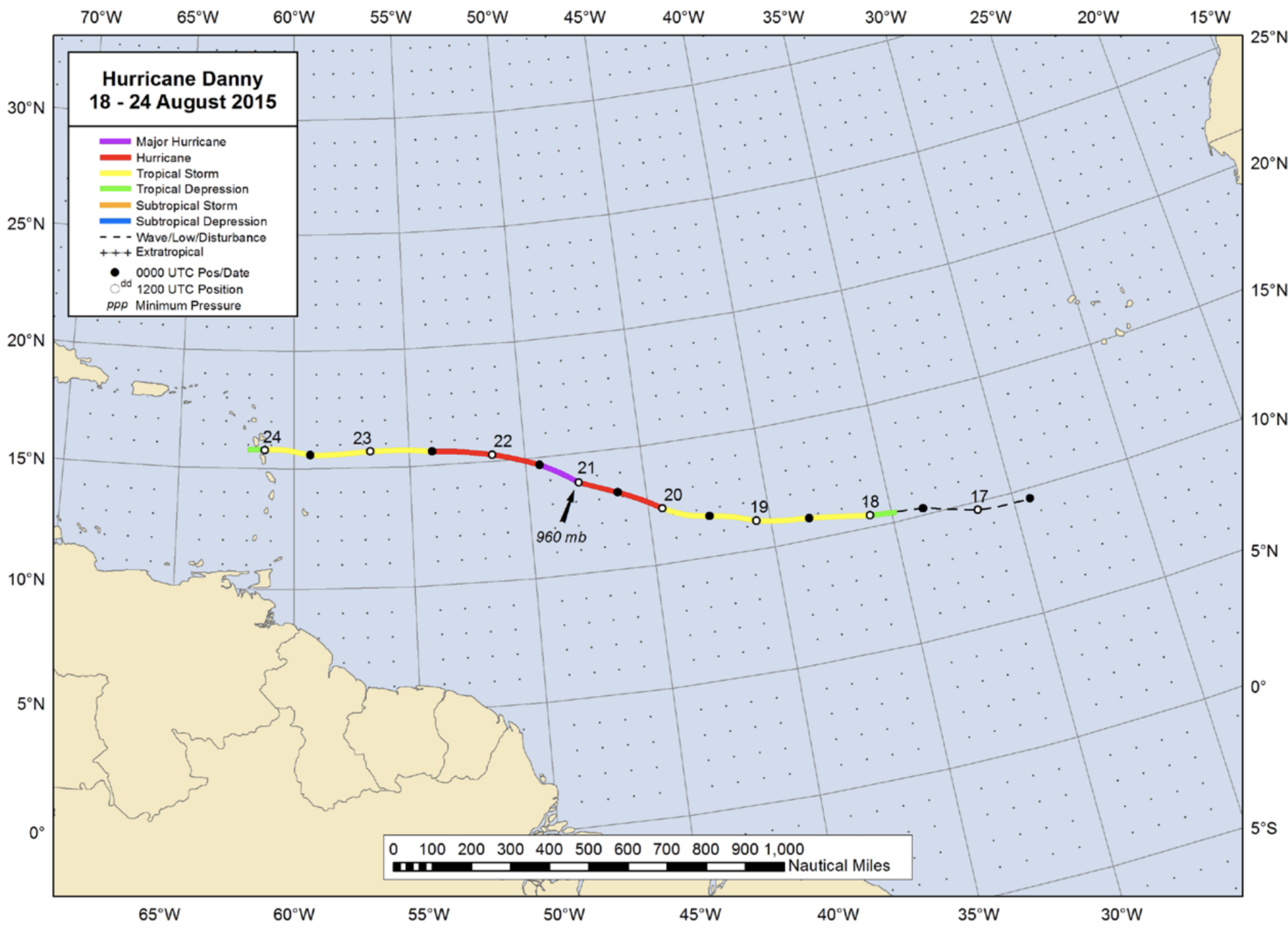}
\caption{NOAA NHC track Hurricane Danny \citep{St2016}.  \label{fig:f1}}
\end{figure}

Hurricane Danny formed as the result of a tropical wave on August 14, 2015. Due to abnormally low environmental wind shear, it relatively quickly reached a minimum pressure of 960 hPa and maximum wind speed of $56 \mathrm{\ m\ s^{-1}}$ at 1200 UTC on August 21 \citep{St2016}. Hurricane Danny was an abnormally small storm, with tropical-storm-force winds extending only to a radius of 60 km from its center. Figure~\ref{fig:f1} shows the storm track for Hurricane Danny from the National Oceanic and Atmospheric Administration National Hurricane Center (NOAA NHC). The domain size we use for action minimization of Hurricane Danny is $3000$ km $\times \ 1800$ km ($500 \times 300$ grid points) and is centered at $15^{\circ}$N and $42^{\circ}$W. 

\begin{figure}[h]
\centering
\includegraphics[width =.85\textwidth]{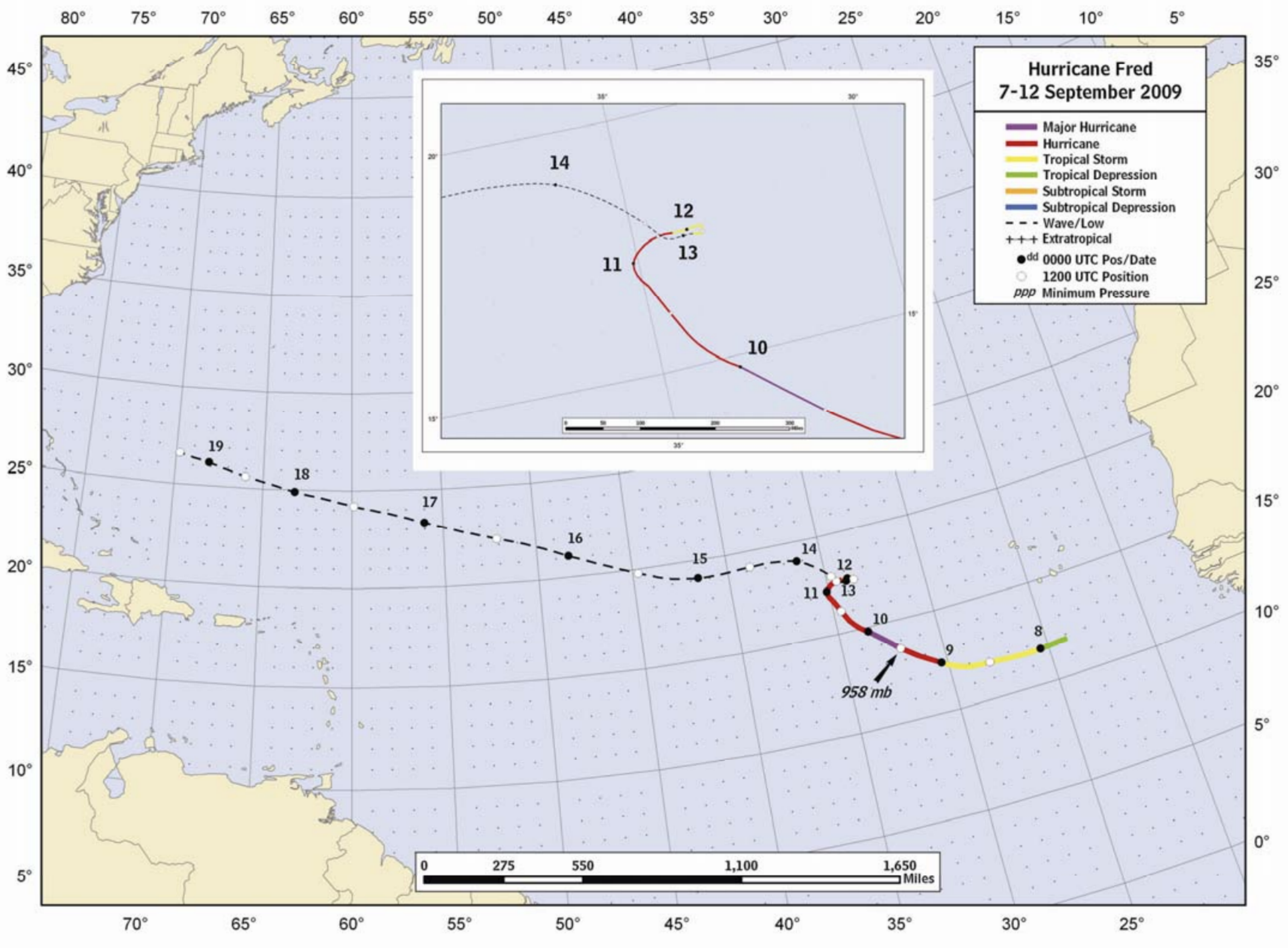}
\caption{NOAA NHC track Hurricane Fred \citep{Br2009}. \label{fig:f2}}
\end{figure}

Hurricane Fred formed from a tropical wave that moved off the west coast of Africa September 6. This depression intensified into Tropical Storm Fred on September 8 and Hurricane Fred on September 9, reaching a minimum pressure of 958 hPa and maximum wind speed of $54 \mathrm{\ m\ s^{-1}}$ later that day \citep{Be2011}. Figure~\ref{fig:f2} shows the storm track for Hurricane Fred from the National Oceanic and Atmospheric Administration National Hurricane Center (NOAA NHC) \citep{Br2009}. The domain size we use for action minimization of Hurricane Fred is $2250$ km $\times \ 2250$ km ($375 \times 375$ grid points) and is centered at $15^{\circ}$N and $28^{\circ}$W. Note that for both Hurricanes Danny and Fred, the domains we use in our simulations are subsets of the domains shown in Figure~\ref{fig:f1} and Figure~\ref{fig:f2}, respectively.

The physics schemes follow the work of \citet{Fi2009} and \citet{Ju2016}. We use the WRF Single-Moment 6-Class Microphysics scheme \citep{Ho2006} to simulate moist processes, the Yonsei University scheme \citep{Ho2006v2} for planetary boundary-layer processes, and the Kain-Fritsch cumulus parameterization scheme \citep{Ka1993}. Ideally, future studies will use higher resolutions without the need for cumulus parameterizations; however, the simulations presented here were limited by machine memory and required 6 km resolution. The parameters used for the action minimization algorithm are: running cost weight $R_i = 0.6$ for each time step $i$ and final cost weight $R_f = 1.0$. These constants are empirically chosen to give maximal weight to the final cost without causing the model to enter unphysical states and crash. Perturbations are added once every hour of integration time, as opposed to at every time step, so as not to incur excessive memory cost associated with storing perturbations at every time step. Low resolution tests suggest that this perturbation frequency is sufficient, as decreasing the perturbation frequency to once every four hours yields nearly identical results to those presented in~\ref{res}. Sensitivity of results to both time horizon and perturbation frequency are discussion in section~\ref{sec:sens}.

In all cases, the minimization of equation~\ref{action functional} is done using a Fletcher-Reeves conjugate gradient method \citep{No2006}. Initial and boundary conditions for all runs are obtained from the NOAA GFS-ANL $0.5^\circ$ dataset.

\subsection{Action minimization for the study of rapid intensification}
\label{sec:justification}

While ensemble simulation is a natural tool for the study of cyclone genesis, intensification, and predictability \citep[e.g.][]{Va2008}, action minimization provides a more computationally efficient way to explore the tail of the cyclone intensity distribution. We test this by comparing the computational cost of using action minimization to cause further intensification of Hurricane Danny with the computational cost of generating an equally intense storm via ensemble simulation. As noted in section~\ref{sec:act}, the background covariance matrix $B$ defined in that section implies a stochastic model wherein the noise is drawn once every hour from a Gaussian distribution with covariance $B$. Action minimization output is then equivalent to a particular realization of this stochastic physical model. We run this model 100 times using the same initial condition ${\bf a}$ as in the action minimization run for Hurricane Danny.

\begin{figure}[h]
\centering
\includegraphics[width =\textwidth]{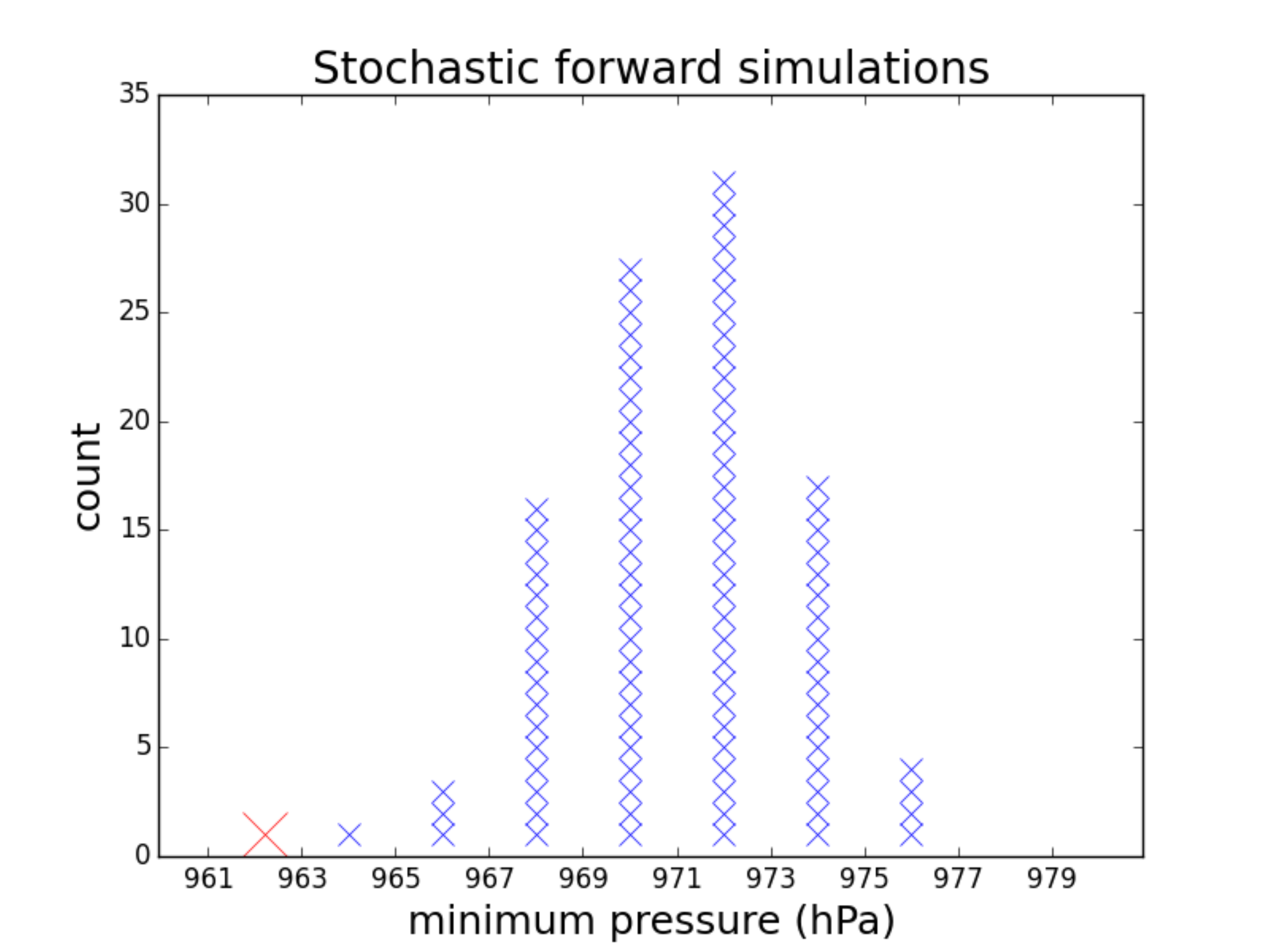}
\caption{Histogram of minimum pressures for Hurricane Danny obtained by ensemble simulation (blue crosses) and the minimum pressure obtained by action minimization (red cross). \label{fig:f3}}
\end{figure}

Figure \ref{fig:f3} shows a histogram of the minimum pressures obtained with the stochastic model and compares them to the minimum pressure (962 hPa) obtained with action minimization. None of the ensemble members reaches as low a minimum pressure as the action minimization output; the closest ensemble member intensifies to 964 hPa. The unperturbed minimum pressure is 970 hPa, while the distribution of ensemble pressure minima has a mean of 971 hPa and a standard deviation of 3.2 hPa. This means that the minimum pressure obtained via action minimization is 2.8 standard deviations lower than the ensemble mean. Assuming a normal distribution of ensemble pressures (this is an approximation since the ensemble distribution appears slightly right-skewed), obtaining an equally low minimum pressure using ensemble simulation is approximately a one in 400 event.

The cost advantage for action minimization relative to ensemble simulation should increase rapidly as one seeks samples further into the tails of the ensemble distribution.  Indeed, if the probability of observing a trajectory $\{{\bf x_i}\}_{i=0}^N$ with $\Phi({\bf x_N})\leq L$ under the stochastic model is $p,$ then the expected number of trajectories required to observe the event $\Phi({\bf x_N})\leq L$ is exactly $1/p.$  For most stochastic models, one would expect that $p$ will decrease very rapidly as $L$ decreases (e.g. exponentially under our Gaussian perturbation model). To assess the efficiency of our implementation of action minimization to access the lower tail of the pressure distribution, we compare the computational cost of running action minimization with the cost of running a 400-member ensemble. The action minimization algorithm uses a conjugate gradient method, which in our experience takes 3 - 7 iterations to converge. Each iteration involves one forward run and one adjoint run to calculate the derivative in equation~\ref{full deriv}, and the cost of each adjoint run is no more than three times the cost of a forward run \citep{Fa2013}. Each iteration also involved a line search in the direction of the gradient, which takes at most 5 forward runs. The computational cost of each iteration is therefore at most 9 times that of a forward run, so the cost of running action minimization is at most equivalent to that of 63 forward runs (operationally, we find that the typical cost is more similar to that of 30 forward runs). Thus, even in the worst case scenario, action minimization is over six times less costly for accessing the tail of TC intensity distribution as compared to ensemble simulation. We expect that there is substantial room to reduce the cost of action minimization by improving upon our initial implementation.

We conclude that while ensemble simulation may be the preferred tool in a number of contexts, the extreme events at the tail of the TC intensification distribution are best accessed using action minimization. Action minimization is thus a complementary tool to ensemble simulation in the sense that the latter should be used to characterize the bulk of the intensification distribution, while the former can be used to more efficiently fill out the tails.

\section{Results}
\label{res}

\subsection{Comparison with maximum potential intensity}
\label{MPI}

It is possible to estimate the maximum potential intensity (MPI) of a TC, usually defined as the minimum theoretically attainable surface pressure. \citet{Em1988} presents a theoretical limit based on the assumption that a TC functions as a Carnot heat engine; \citet{Ho1997} presents a thermodynamic limit based on empirical relationships among relative humidity, central pressure, and wind speed. Most TCs do not attain maximum potential intensity, with the typical pressure depression magnitude approximately 55\% of that predicted by MPI \citep{De1994}. The reasons for TCs failing to attain MPI could include dynamical controls, e.g. environmental flow and vertical shear \citep{Ze2007}. It is therefore natural to ask whether action minimization can modify the dynamics in such a way as to bring TCs closer to MPI.

\begin{figure}[h]\center{
\includegraphics[width = \textwidth]{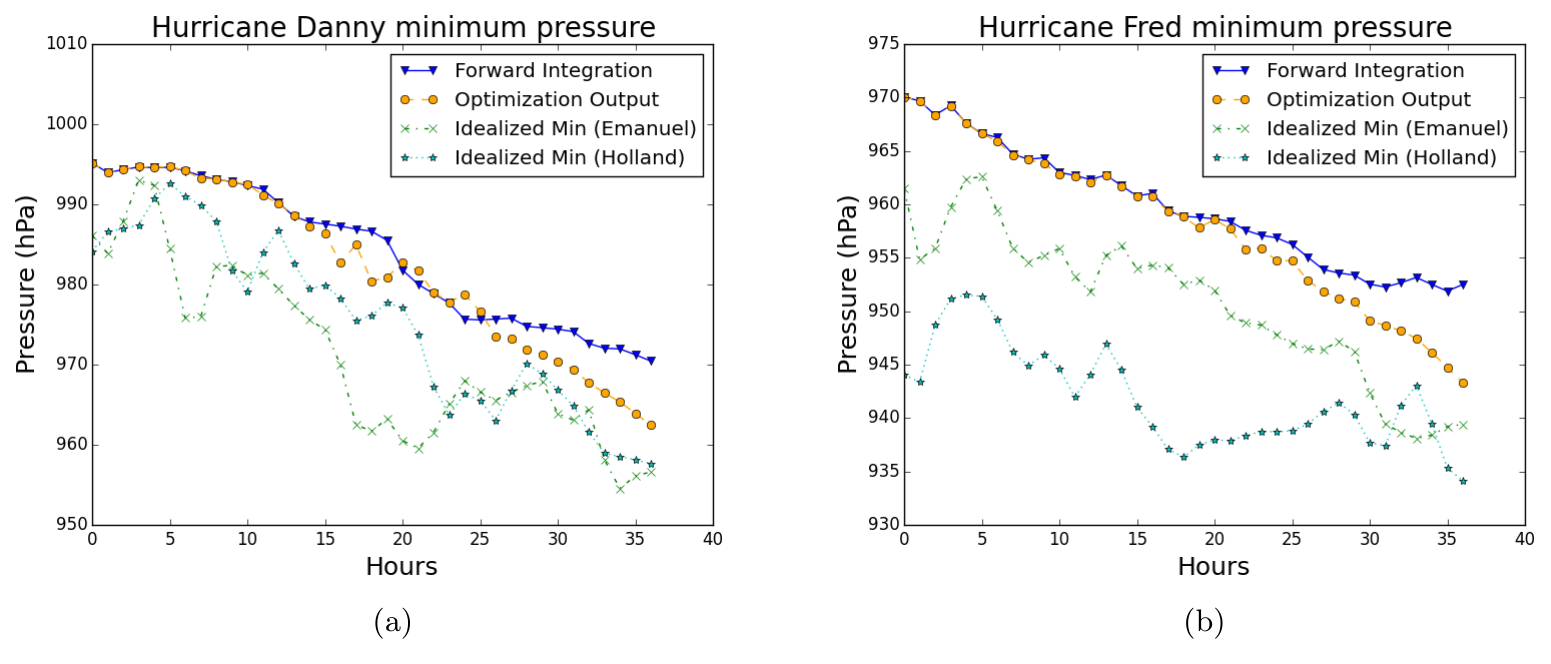}
\caption{Minimum pressure as a function of time for forward integration (blue triangles) and action minimization output (orange circles) of Hurricanes Danny (panel a) and Fred (panel b). The green crosses and cyan stars are maximum potential intensities according to \citet{Em1991} and \citet{Ho1997}, respectively.}
\label{fig:f4} }
\end{figure}

Figure~\ref{fig:f4} shows a comparison of the minimum pressures of the action minimization output and simple forward integration, as well as Emanuel and Holland MPI for Hurricanes Danny and Fred. We obtain the Emanuel MPI by solving equation 16 in \citet{Em1991}, which is an implicit relationship for the central pressure derived from balancing the work done by a TC (to drive surface wind) against frictional dissipation in the boundary layer. A correction for dissipative heating is done according to \citet{Bi1998}. Solving this equation requires quantities easily obtained from WRF output, including temperature at the sea surface and at the top of the troposphere, ambient sea level pressure, and relative humidity at both the center and outer radius of the storm. We obtain Holland MPI by substituting into equation 2 of \citet{Ho1997}, which requires the same quantities.

\begin{figure}[h]\center{
\includegraphics[width = \textwidth]{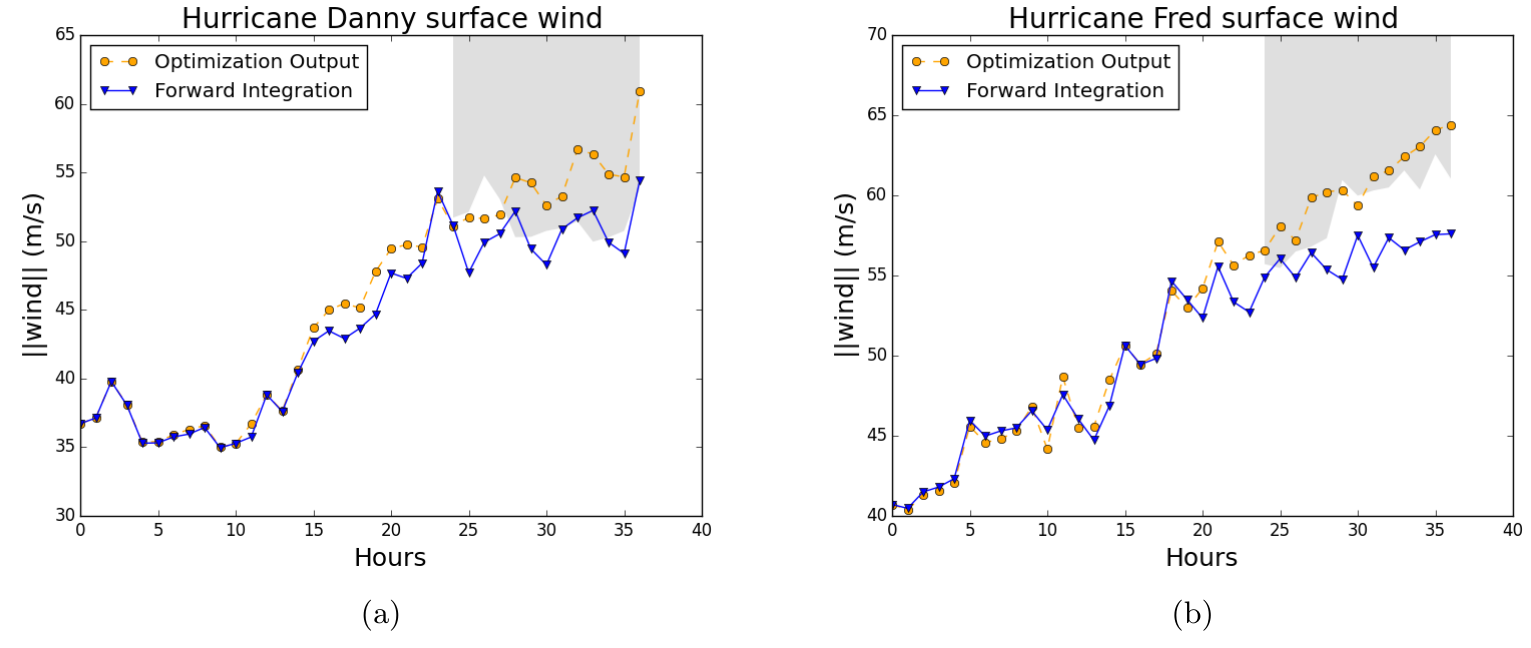}
\caption{Maximum surface winds as a function of time for forward integration (blue triangles) and action minimization output (orange circles) of Hurricanes Danny (panel a) and Fred (panel b). The shaded regions mark wind speeds that must be reached for RI to occur. }
\label{fig:f5} }
\end{figure}

The optimization for Hurricane Danny results in a decrease in minimum central pressure from 970.5 hPa to 962.0 hPa, accompanied by an increase in maximum surface wind speed from from 54.4 m s$^{-1}$ to 60.9 m s$^{-1}$ (Fig~\ref{fig:f5}). For Hurricane Fred, the pressure decreases from 953.1 hPa to 943.3 hPa and the maximum wind speed increases from 57.6 m s$^{-1}$ to 64.4 m s$^{-1}$. Thus, action minimization causes both storms to intensify from category three to category four. Figure~\ref{fig:f5} shows the maximum surface winds as a function of time. The shaded regions show how fast surface wind must be to meet the RI criterion of intensification by at least $15$ m s$^{-1}$ over 24 hours. The unperturbed forward integrations of both Hurricanes Danny and Fred are on the borderline of the RI criterion; the action minimization output clearly meets the RI criterion during the last several hours of both cases. The action minimization algorithm is therefore able to ensure that RI occurs for these storms.

Although the optimization runs show similar decreases in pressure and increases in wind speed, the nature of the divergence of the optimized run from the forward integration is not the same in the two cases. While both optimization runs diverge from the forward integration during the second half of the time horizon, the optimized pressure trajectory for Hurricane Danny falls slightly above the pressure trajectory of the forward integration at several points. This is in contrast with the optimized pressure trajectory for Hurricane Fred, which remains below the forward integration trajectory once the two begin to diverge. We speculate that this is a result of differences in heating patterns discussed in section~\ref{sec:T}.

Importantly, neither run results in lower central pressure than that indicated by maximum potential intensities, suggesting that the more intense storms resulting from action minimization are thermodynamically feasible. We verify this by integrating both storms forward with no action minimization from both the perturbed and unperturbed states at the final time in Figure~\ref{fig:f4}. In both cases, the central pressures rise as the storms pass peak intensity, with those from the perturbed runs eventually approaching those from the forward integration. Further, we conduct tests at lower resolution in which we restart the optimization after 36 hours; that is, we use the final state of the original action minimization output as the initial state of a new action minimization problem with the subsequent 36 hours as its time horizon. This yields storms that stay near MPI for a longer time, but even in these simulations the minimum central pressures do not cross the MPI by more than 1-3 hPa or for longer than one optimization time step. 

Importantly, MPI should not be interpreted as a hard bound on TC intensity, as there are occasional cases of TC intensity exceeding MPI (e.g. Hurricane Patricia \citep{Ro2017}). While comparison with MPI alone does not ensure that our results are physically realistic, it provides a useful estimate of what is thermodynamically feasible and can be used to check that the results presented here are within reasonable bounds. Notably, when we try to generate storms that significantly exceed MPI (by decreasing $R_f$ and therefore increasing the importance of the final cost), the perturbations become too large and cause the model to crash. In sum, our results suggest that the perturbed simulations are broadly consistent with thermodynamic constraints on TC intensity.

\subsection{Comparison of perturbed and unperturbed storm structure}

\begin{figure}[h!]\center{
\includegraphics[width = .95\textwidth]{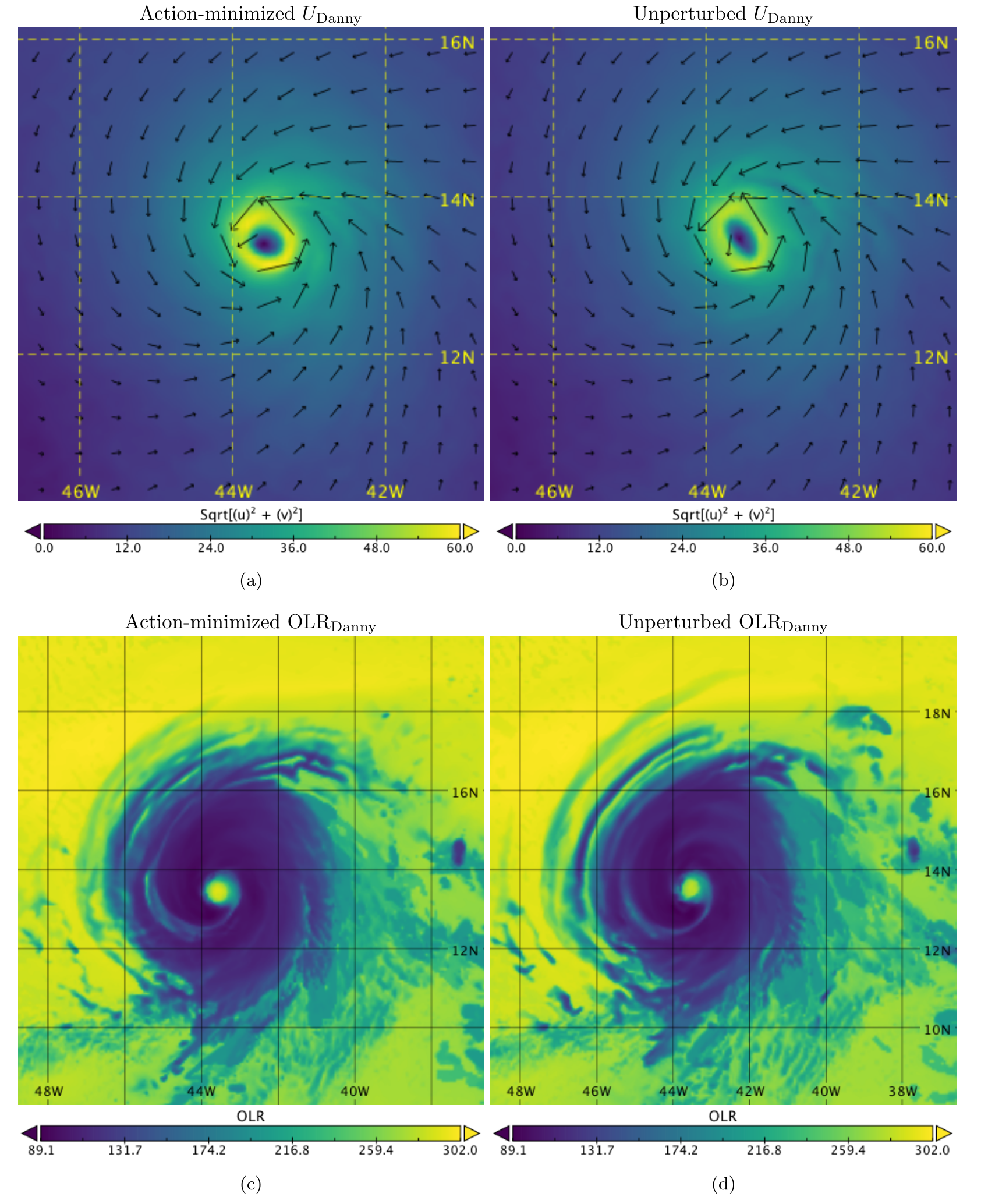}
\caption{Surface wind magnitude and OLR snapshots at final integration time step for Hurricane Danny: optimization output (panels a and c) and forward integration (panels b and d). }
\label{fig:f6} }
\end{figure}

Figure~\ref{fig:f6} shows contour plots of the surface wind and the outgoing longwave radiation (OLR) at the final time for the forward integration and the optimization output for Hurricane Danny. Both fields show no qualitative structural inconsistencies between the forward integration and the optimization output. In particular, the optimization contains the key expected features including rainbands and an eyewall. The radius of maximum wind decreases from approximately 38 km in the unperturbed forward integration to approximately 36 km in the action minimization output. Both surface wind and OLR fields for Hurricane Fred (not shown) exhibit similar behavior, with action minimization once again causing a slight reduction in the radius of maximum wind but no major qualitative changes. 

\subsection{Temperature perturbations}
\label{sec:T}

\begin{figure}[h!]\center{
\includegraphics[width = .705\textwidth]{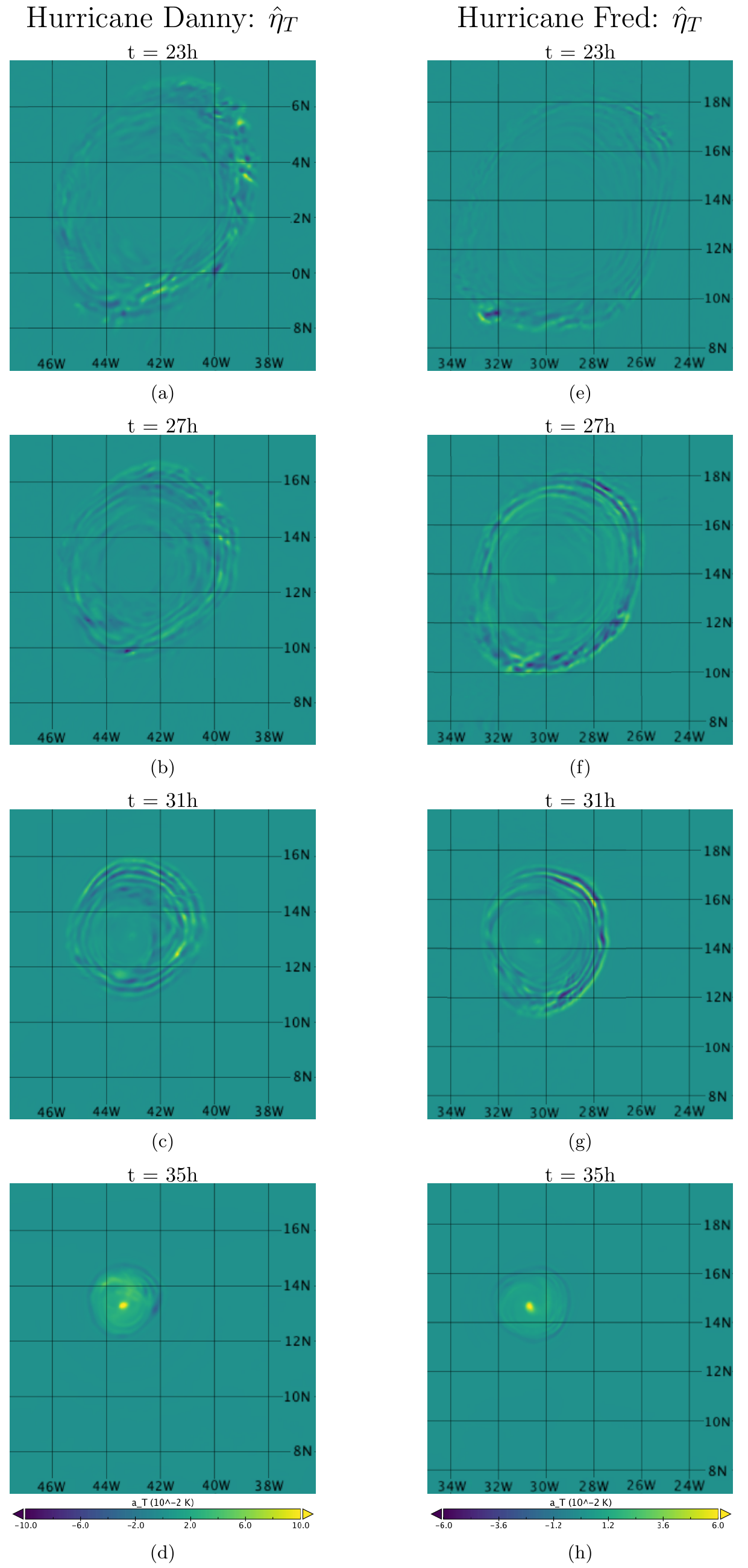}
\caption{Temperature perturbations ${\bf \hat{\eta}_T}$ at 900 hPa height for action minimization of Hurricanes Danny (panels a-d) and Fred (panels e-h).}
\label{fig:f7}}
\end{figure}

To determine the mechanisms responsible for intensifying the optimized storm over the forward integration, we investigate the perturbations ${\bf \hat{\eta}_i}$ added to the forward trajectory. Figure~\ref{fig:f7} shows the 850 hPa temperature perturbations ${\bf \hat{\eta}_T}$ added 13, 9, 5, and 1 hours before the minimum pressure is attained for both Hurricane Danny (panels a-d) and Hurricane Fred (panels e-h). 

\begin{figure}[h!]\center{
\includegraphics[width = .705\textwidth]{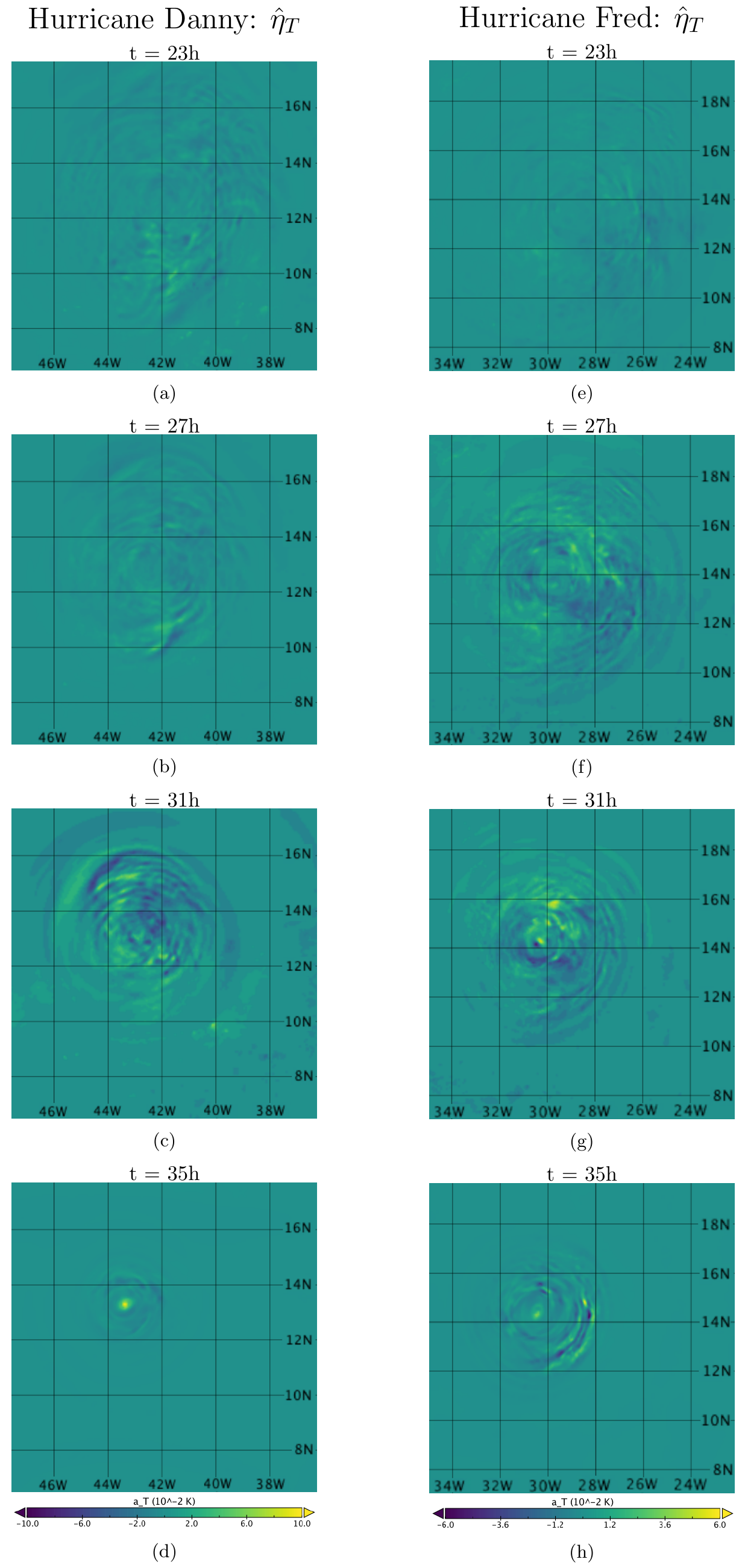}
\caption{As in Figure~\ref{fig:f7}, but at 500 hPa instead of 850 hPa.}
\label{fig:f8}}
\end{figure}

Many previous studies have demonstrated the importance of warm cores in TC intensification \citep{Sc1982, Si1997}, so one may expect action minimization to perturb temperature in such a way as to directly strengthen the warm cores of Hurricanes Danny and Fred. Instead, the algorithm adds rings of low level warm temperature to the outside of the developing storms. The ring radius is initially approximately 250 km for Hurricane Danny and 350 km for Hurricane Fred, which is consistent with the radii of the two storms. As the TCs intensify, the warm air at the outside of the storms moves toward the storm centers. This warm air reaches radii of 40 km and 60 km radii for Danny and Fred, respectively, thereby strengthening the existing warm cores. This is consistent with~\citet{St2012}, who used the WRF model to demonstrate a warm core horizontal length scale of $\mathcal{O}(50 \mathrm{ km})$ in a mature storm. In both cases, the entirety of the ring of warm air consistently penetrates up to $650$ hPa ($z \approx 3$ km) with maxima in the $800-850$ hPa range, but also contains regions that penetrate up to $400$ hPa toward the end of the simulations (Figure~\ref{fig:f8}). 

\begin{figure}[h]
\includegraphics[width = \textwidth]{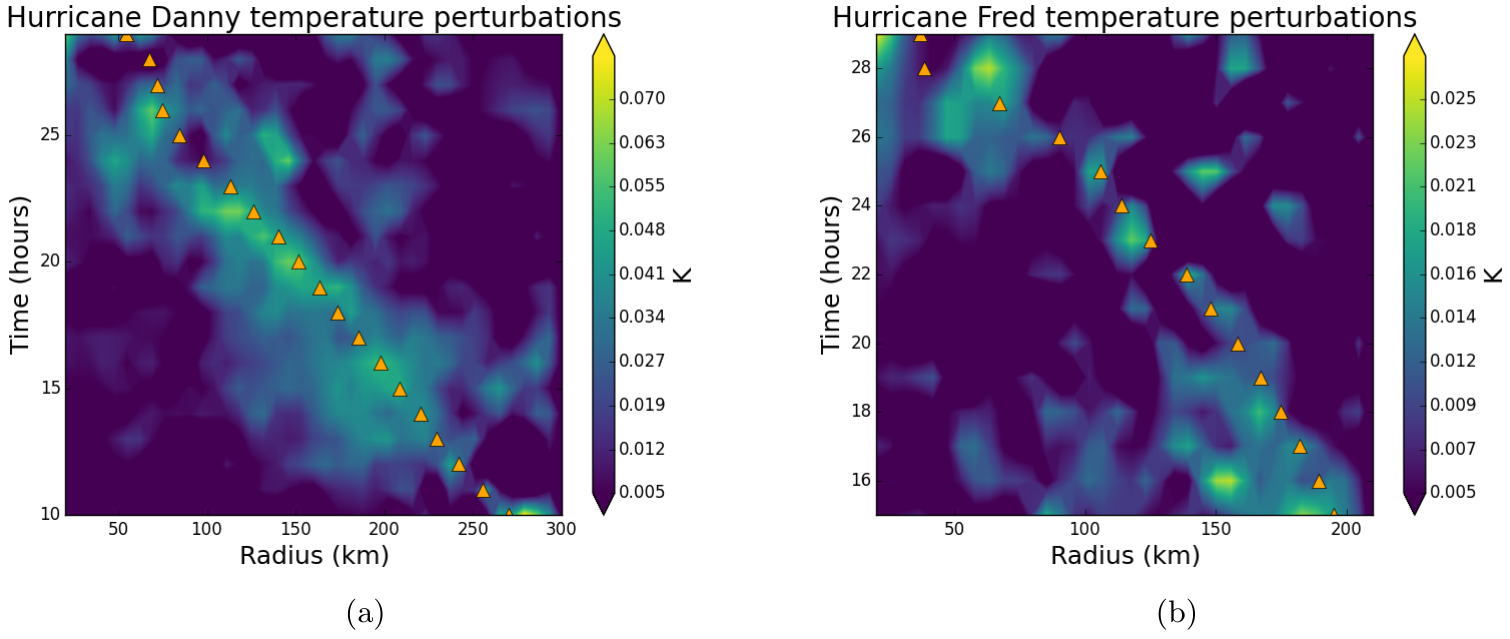}
\caption{Hovm\"{o}ller diagrams of 900 hPa temperature perturbations with radius and time for Hurricanes Danny and Fred. Orange triangles indicate the position of an air parcel advected by mean radial flow.}
\label{fig:f9}
\end{figure}

Interestingly, the temperature perturbations are not exclusively positive and exhibit a wave-like pattern in the radial direction. This is somewhat counterintuitive, since one may expect that action minimization would choose exclusively to add energy to the system. It is possible that the negative perturbations serve to contribute to the eventual axisymmetrization of the storm at the end of the time horizon; however, this is speculative and more work is required to understand why negative perturbations are chosen. It is important to note that, when integrated over the area of the storm, the positive temperature perturbations dominate and the total temperature contribution of action minimization is positive. Further, as we show in Figure~\ref{fig:f9}, perturbations move in a way that is consistent with advection rather than with gravity waves.

While individual perturbations are instantaneous and so cannot be advected (once a perturbation becomes part of the state, it is no longer possible to distinguish it from the underlying state at further times), we hypothesize that the temperature perturbations are consistent with advection toward the center of the storm by converging boundary layer air in the following sense: 
\begin{enumerate}
\item{ A perturbation is added at time $t_i$ at radius $r_i$.} 
\item{ Air parcels at $t_i$, $r_i$ are advected to a new radius $r_{i+1}$ at time $t_{i+1}$.} 
\item{ The perturbation at time $t_{i+1}$ is added at this new radius $r_{i+1}$.} 
\end{enumerate}

In the Lagrangian perspective, the algorithm heats the same parcels of air throughout the time horizon of the simulation; specifically, it heats those parcels that reach the radius of maximum wind at the end of the simulation. Figure~\ref{fig:f9} shows Hovm\"{o}ller diagrams of 900 hPa ($z \approx 1.5$ km)  temperature perturbations with radius and time for Hurricanes Danny and Fred. The orange triangles indicate the positions of parcels of air being advected by the radial flow (calculated by integrating the velocity field). In both cases, the alignment between the trajectory of the perturbations and the trajectory of the advected parcels shows that action minimization is choosing to heat roughly the same parcels as they are advected inward toward the center of the storm. During the last five hours of each run, temperature perturbations are dominated by heating near the center of the storm (not shown in Figure~\ref{fig:f9} for clarity of the color bar, but apparent in panels d and h of Figure~\ref{fig:f7}). Thus, even though heating during the early stages of the storm may be relatively inefficient since it occurs outside of the radius of maximum wind \citep{Vi2009}, the action minimization algorithm finds a path to transport this heat to the center of the storm and efficiently cause intensification toward the end of the simulation. This may explain why the pressure trajectories shown in Figure~\ref{fig:f4} show larger divergence from the unperturbed runs during the later stages of the action minimization output.

Numerous studies have investigated the roles of symmetric and asymmetric heating in TC intensification \citep{Ca1989, No1999, No2002, No2003, No2007, Sh2000, Mo2013}. By decomposing temperature perturbations into an azimuthally averaged symmetric component and a mean-zero asymmetric component, \citet{No2003} found that the response of a TC wind field can be closely approximated by the purely symmetric response to the symmetric component of the heating. \citet{No2007b} confirmed this result, but did find that there are times when vortices with relatively small radii can be strengthened by the asymmetric heating.

While these studies used some temperature and velocity perturbations with time-varying azimuthal structure, the radial structure of the perturbations was constant with time. Further, \citet{No2007b} used temperature perturbations with maxima at $z  = 6$ km and that decayed to 0 at $z = 3$ km. Such perturbations are consistent with some observational studies of latent heating distributions \citep{ya2002} and with modeling studies in which deep convection is the source of heating \citep{He2004, Mo2006}. However, both observational and modeling studies suggest that latent heating in certain storms can be significant at much lower altitudes \citep{Pa2013, Za2014b, Ta2015}. \citet{Mo2005} found that, depending on the nature of pre-existing asymmetries in a vortex, asymmetric heating can cause TC intensification of a similar magnitude (both in pressure and maximum wind) and over a similar time horizon as in this study. Importantly, while there are varied and sometimes contradicting studies regarding the impact of asymmetry and asymmetric heating on TC intensification, these studies all use prescribed perturbations to the temperature and/or PV fields; their conclusions are thus limited to the particular perturbations under consideration. By searching over the full space of possible temperature perturbations with action minimization, this study advances this line of inquiry.  

\begin{figure}[h]
\includegraphics[width = \textwidth]{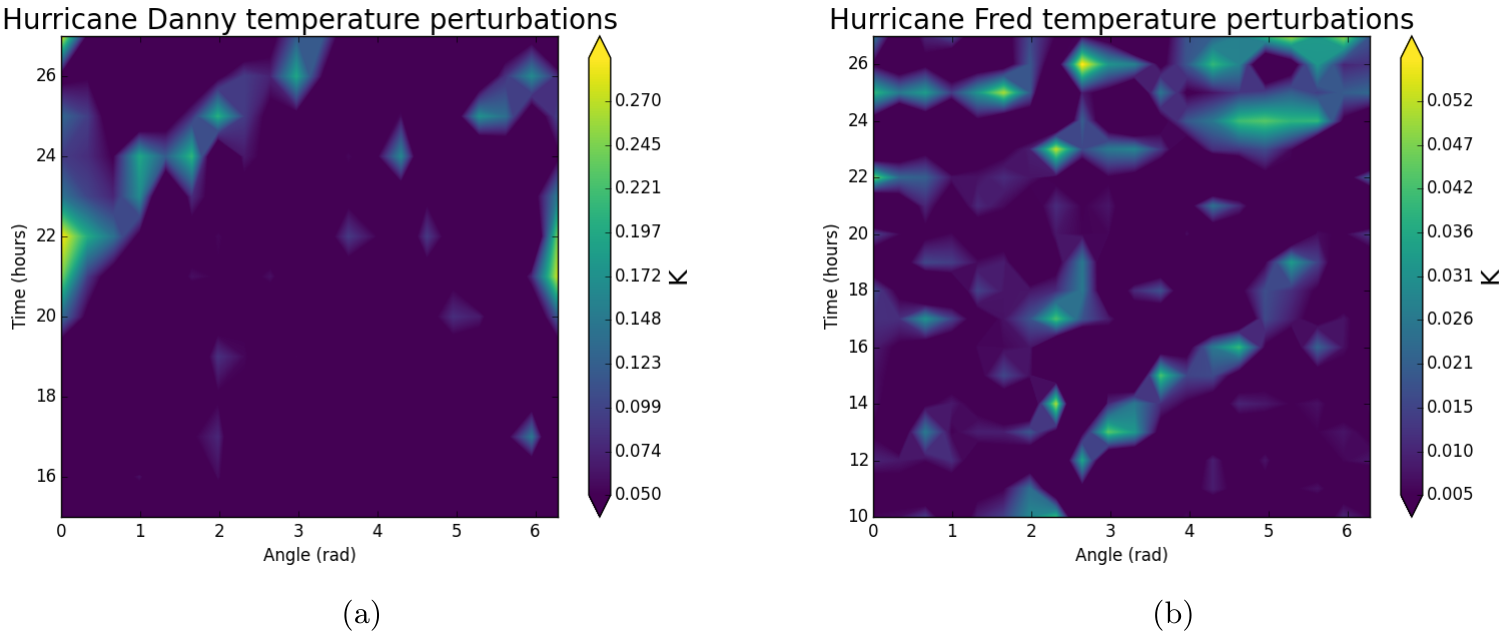}
\caption{Hovm\"{o}ller diagrams of 900 hPa temperature perturbations, averaged over $r \in [50, 150]$ km, with angle and time for Hurricanes Danny and Fred.}
\label{fig:f10}
\end{figure}

Figure~\ref{fig:f10} shows Hovm\"{o}ller diagrams of 900 hPa temperature perturbations, averaged over $r \in [50, 150]$ km, with angle (relative to the center of the storm, with $\theta = 0$ being directly east of the storm and increasing cyclonically) and time for Hurricanes Danny (panel a) and Fred (panel b). Perturbations for Hurricane Danny appear to take on an asymmetric structure 20 hours into the simulation, as suggested by the fact that they are limited to a single, small range of angles after that time. This contrasts sharply with perturbations for Hurricane Fred, which show little coherent dependence on angle.

\begin{figure}[h]
\includegraphics[width = \textwidth]{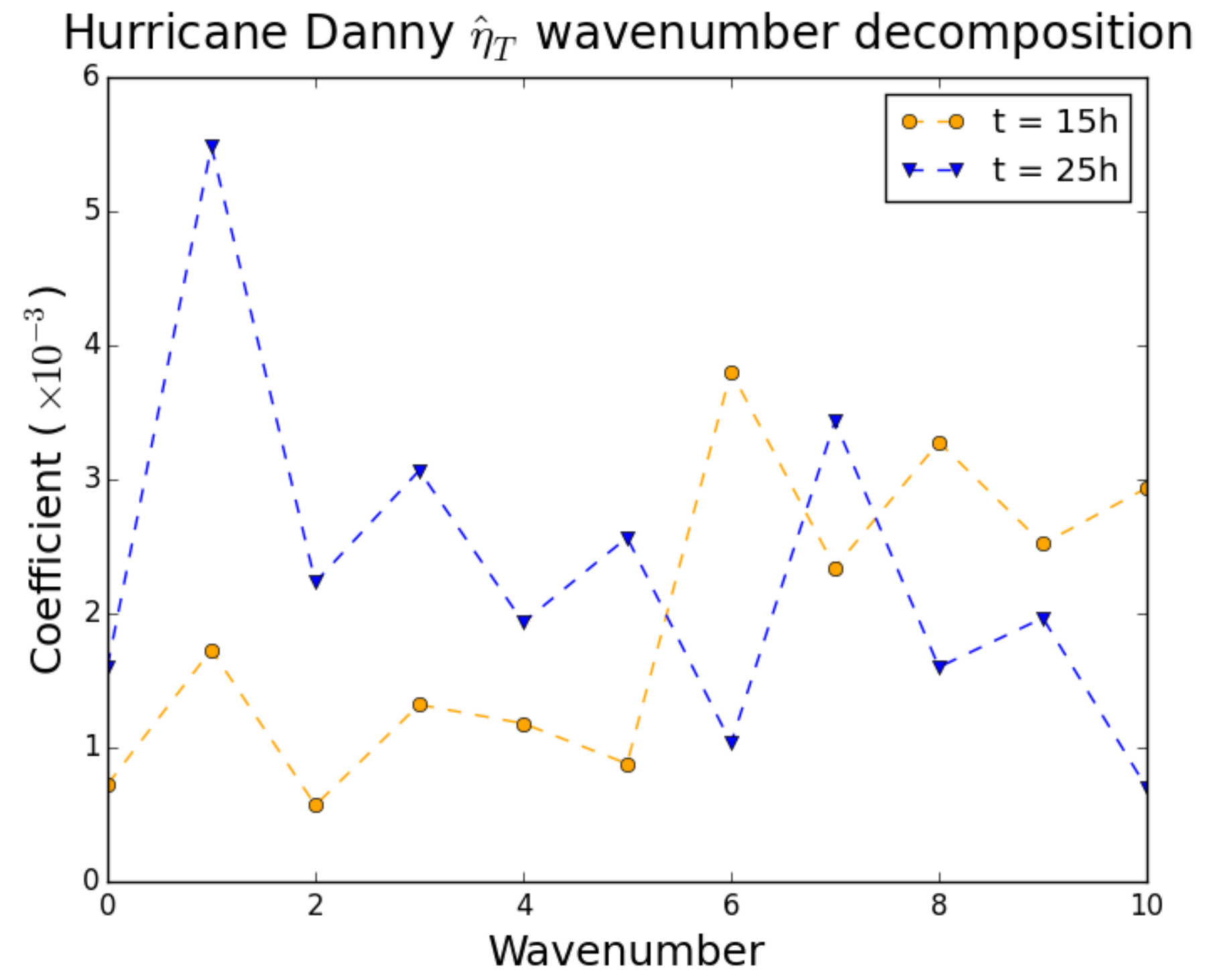}
\caption{{Wavenumber decomposition of Hurricane Danny temperature perturbations at $t = 15$h (yellow curve) and $t = 25$h (blue curve).}}
\label{fig:f11}
\end{figure}

By using Fourier decomposition of the temperature perturbation field at a given radius, we find that temperature perturbations for Hurricane Danny are dominated by relatively high wavenumbers ($n \ge 6$) early in the time horizon. This contrasts with the previous studies, including~\citet{No2003} and~\citet{No2007}, who studied the role of asymmetry using low-wavenumber perturbations ($n \leq 4$). However, after $t \approx 18$ hours, wavenumber-one perturbations become relatively more important than perturbations at all other wavenumbers. This behavior is apparent in Figure~\ref{fig:f11}, which shows the wavenumber decomposition of the temperature perturbations to Hurricane Danny at $t=15$h and $t=25$h The different azimuthal structure of the asymmetries imposed by action minimization as compared to those used in previous studies may impact our finding that asymmetry can contribute to intensification. This, and the evolving nature of the action-minimizing asymmetries over time, bears further study.

\begin{figure}[h]
\includegraphics[width = \textwidth]{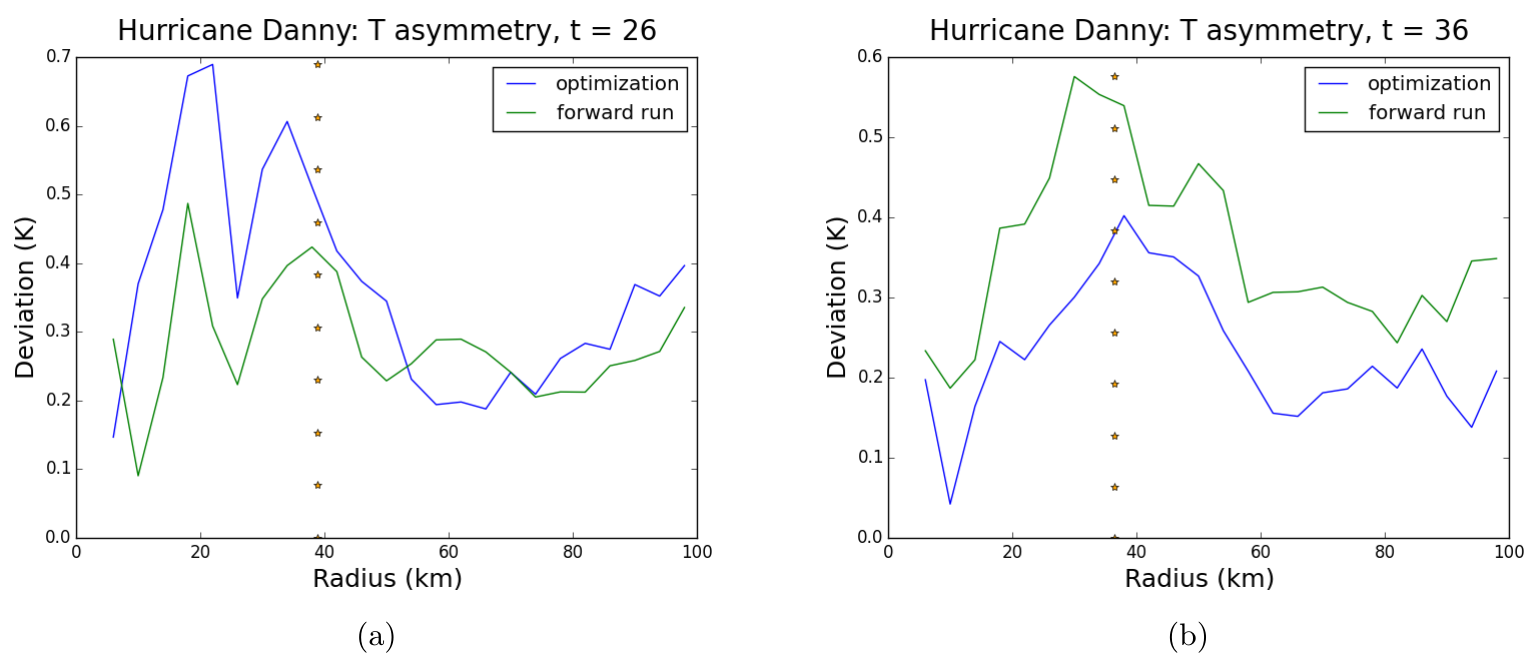}
\caption{Average deviation from symmetry of the 900 hPa temperature field for Hurricane Danny at $t=26$ hours (panel a) and $t=36$ hours (panel b). Orange stars mark the radius of maximum wind.}
\label{fig:f12}
\end{figure}

To test the significance of the asymmetric component of the temperature perturbations to Hurricane Danny, we first establish that the asymmetry of the perturbations does not serve simply to symmetrize the temperature field. We do so by decomposing the 900 hPa temperature field into symmetric and asymmetric components and then calculating the average magnitude of the deviation from symmetry as a function of radius from the center of the storm. Figure~\ref{fig:f12} shows these deviations from symmetry for both the action minimization output and the unperturbed forward run for Hurricane Danny at $t=26$ hours (panel a) and $t=36$ hours (panel b). It shows while that the temperature field becomes more symmetric in the optimization at the final time, it is less symmetric near the radius of maximum wind as compared to the forward run at prior times. The relative asymmetry of the action minimization output is strongest between $t=20$ hours and $t=28$ hours; notably, this corresponds to the time at which the minimum pressure in the optimization output is higher than that of the forward integration. 

\begin{figure}[h!]
\includegraphics[width = .95\textwidth]{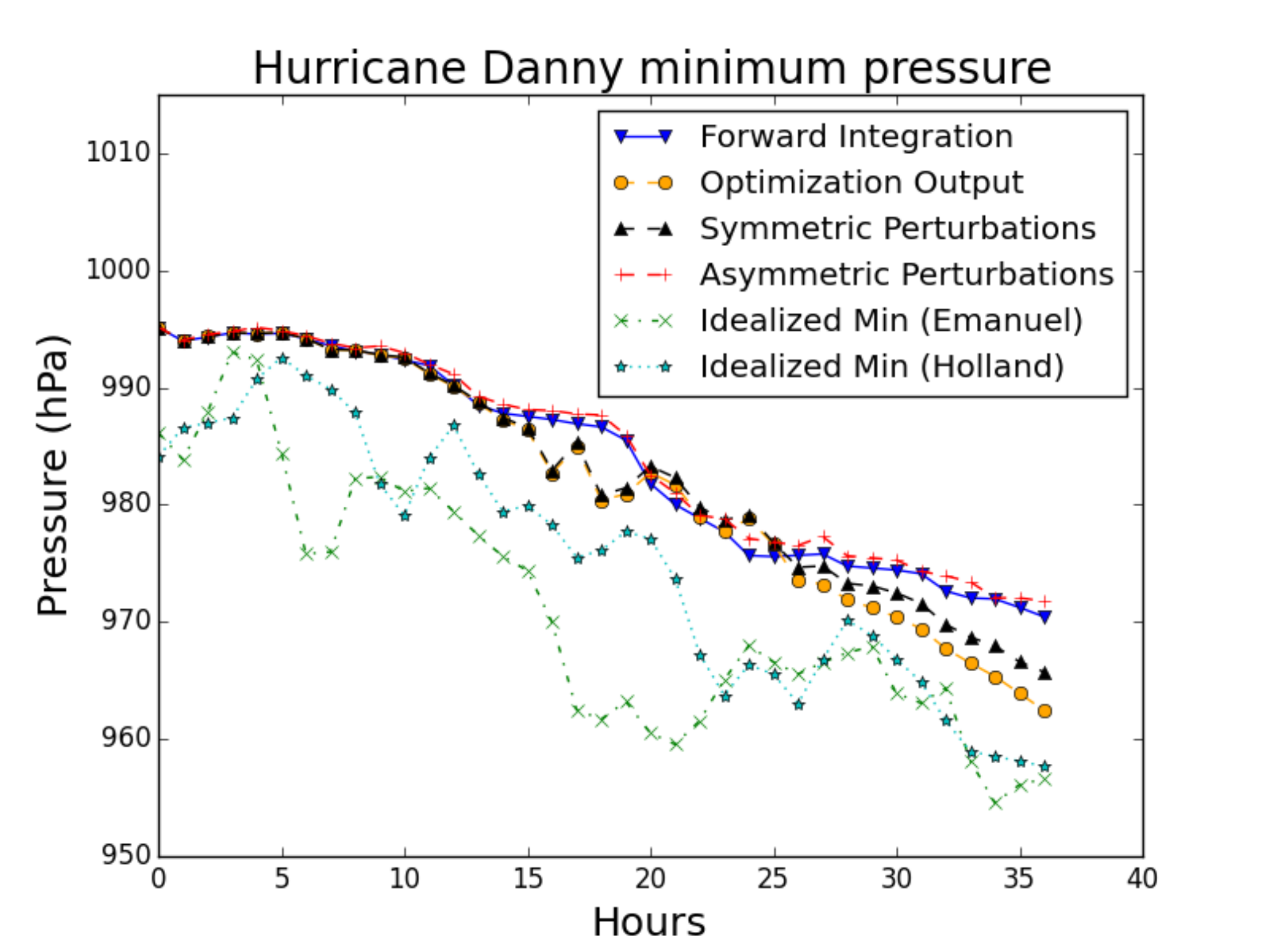}
\caption{As in Figure~\ref{fig:f4}a, but with the purely symmetric of portion of action-minimizing perturbations (black triangles) and with purely asymmetric portion of the perturbations (red crosses) for Hurricane Danny.}
\label{fig:f13}
\end{figure}

This suggests that the asymmetric component of the temperature perturbations may contribute to intensification in the action minimization output. To test this, we run a forward model and add only the symmetric component of the temperature perturbations to the temperature field at each time step. Figure~\ref{fig:f13} shows that, for Hurricane Danny, this configuration yields a minimum pressure of 964.5 hPa. Given that the full action minimization output and unperturbed run yielded minimum pressures of 962 hPa and 970.5 hPa, respectively, removing the asymmetry results in a $30\%$ reduction in intensification. Notably, repeating this test for Hurricane Fred results in no significant change to the minimum pressure relative to the optimization output (not shown). While the difference between runs with and without asymmetry of perturbations is quite small, the fact that it comprises $30\%$ of the additional intensification achieved through action minimization suggests that it could be meaningful. A larger sample size of storms where the action-minimizing perturbations are noticeably asymmetric will be necessary to confirm the significance of the asymmetry. We also check the effect of the purely asymmetric component of the action-minimizing perturbations; Figure~\ref{fig:f13} shows that forward integration with purely asymmetric perturbations yields a small ($< 1$ hPa) weakening of Hurricane Danny. This is consistent with \citet{No2003} and \citet{No2007}. However, the fact that the linear combination of the symmetric and asymmetric perturbations produces stronger intensification than the symmetric perturbations alone suggests a significant nonlinearity in storm response to perturbations that was not observed in previous studies.

\subsection{Velocity response to heating}

As with temperature, we decompose the radial and azimuthal wind fields for Hurricanes Danny and Fred into symmetric and asymmetric components. Figure~\ref{fig:f14} shows the average of the magnitude of deviations from symmetry of the azimuthal velocities for both storms as a function of radius. Both storms show more symmetric azimuthal wind fields in the optimization output than in the forward integration. The maximum average asymmetry at the final time occurs near the radius of maximum wind for both storms, which is approximately 30 km for Danny and 50 km for Fred. The deviations from symmetry of the radial wind fields (not shown) have very similar responses to those of the azimuthal fields.

\begin{figure}[h!]
\includegraphics[width = .95\textwidth]{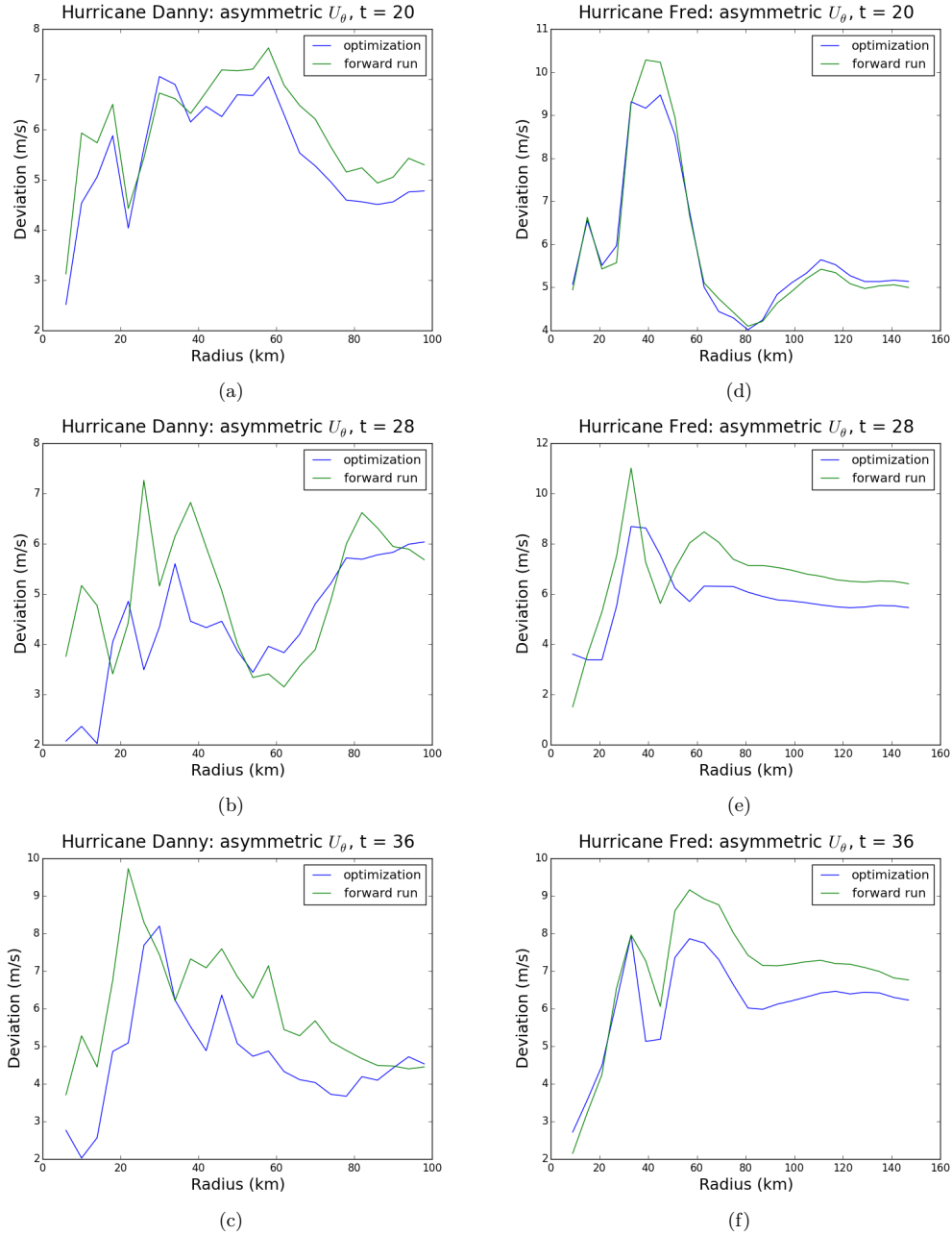}
\caption{Average deviation from symmetry of the azimuthal surface wind field as a function of radius for Hurricane Danny (panels a-c) and Hurricane Fred (panels d-f) at $t=20,\ 28,\ \mathrm{and\ } 36$ hours.}
\label{fig:f14}
\end{figure}

Previous studies have also shown that vertical wind shear is a crucial factor inhibiting TC intensification \citep{Me1988}. \citet{De1996} notes that the effect of vertical shear is to advect upper level heat and moisture away from the low level circulation, thereby disrupting the vertical structure and suppressing storm development. \citet{Jo1995} argues that vertical shear displaces upper-level potential vorticity (PV) relative to low level PV, which inhibits convection and intensification. \citet{Ri2010} and \citet{Ta2010} argue that vertical wind shear ventilates mid-level storm cores and introduces cold, dry air, as well as reducing the thermodynamic efficiency of TCs.

\begin{figure}[h]
\includegraphics[width = \textwidth]{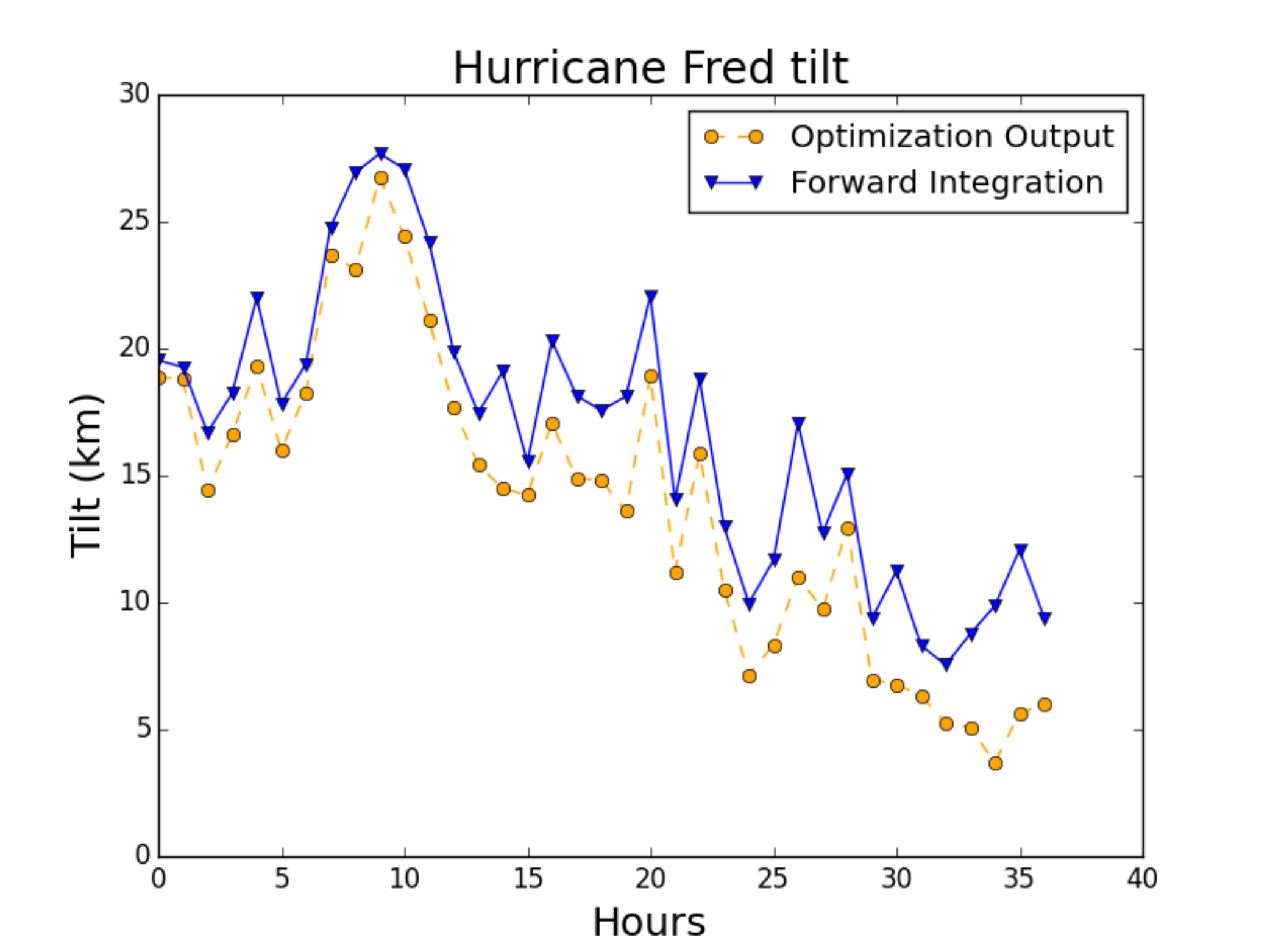}
\caption{Magnitude of 900 - 500 hPa and 850 - 250 hPa wind shear as a function of time for Hurricanes Danny (panels a and b) and Fred (panels c and d) for forward integration (blue curve) and optimization output (orange curve). Markers denote times at which perturbations are added by action minimization.}
\label{fig:f16}
\end{figure}

Following \citet{Ra2012} and \citet{Zh2013}, we define low-level vertical shear as the difference between horizontal winds at 900- and 500-hPa levels averaged over a circle marking the edge of the storm, where the edge is the radius of $8 \mathrm{\ m\ s}^{-1}$ wind \citep{Sc2017}; we define high-level shear analogously but using the 850- and 250-hPa levels \citep{Fr2001, Co2003}. Figure~\ref{fig:f15} shows time series of both low- and high-level vertical shear in the action minimization output and unperturbed forward integration for Hurricanes Danny and Fred. Hurricane Fred shows a noticeable reduction in low-level shear in the action minimization output, with the strongest effect toward the beginning of the run. In contrast, the shear reduction for Hurricane Danny is relatively minor and may well be insignificant. Initial time shear is reduced from $6.11 \mathrm{\ m\ s}^{-1}$ to $5.08 \mathrm{\ m\ s}^{-1}$ for Hurricane Fred and from $3.79 \mathrm{\ m\ s}^{-1}$ to $3.60 \mathrm{\ m\ s}^{-1}$ for Hurricane Danny. This is consistent with~\citet{Wo2004}, who found that TC intensity is most sensitive to changes in shear above a threshold that lies between $2-4 \mathrm{\ m\ s}^{-1}$. Hurricane Danny, which formed in an abnormally low-shear environment, is already at or near this threshold in the unperturbed run, so the algorithm does not waste energy (running cost) reducing its shear. Further, the action-minimizing perturbations reduce the shear as defined by the difference between 900- and 500-hPa level velocities, but not as defined by the difference between 850- and 200-hPa level velocities; this indicates that the low-level shear may be more important for TC intensification in WRF. Consistent with \citet{Fi2016}, the effect of the reduction in low-level shear for Hurricane Fred is to reduce the tilt, as defined by the difference between the location of the storm center at the 900- and 500-hPa levels (calculated using weighted horizontal circulation storm centers following \citep{Zh2014}). Figure~\ref{fig:f16} shows the time evolution of the tilt magnitude for Hurricane Fred; the tilt at the final time is reduced from 9.6 km to 6.1 km in the action minimization output.

\begin{figure}[h]
\includegraphics[width = \textwidth]{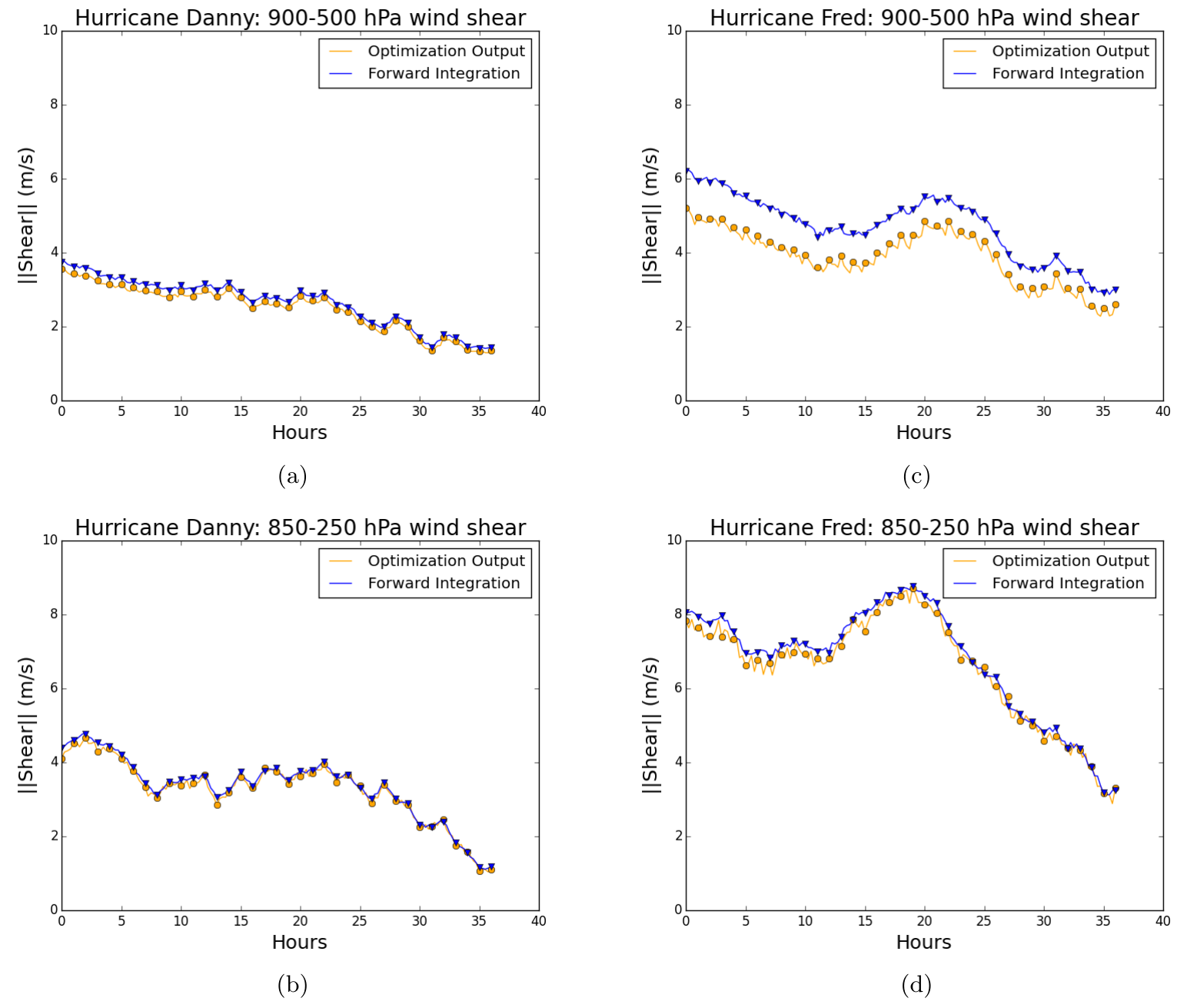}
\caption{Magnitude of 900 - 500 hPa and 850 - 250 hPa wind shear as a function of time for Hurricanes Danny (panels a and b) and Fred (panels c and d) for forward integration (blue curve) and optimization output (orange curve). Markers denote times at which perturbations are added by action minimization.}
\label{fig:f14}
\end{figure}

Finally, we check that the reduction in wind shear does not immediately dissipate after the perturbations are added by calculating the magnitude of the shear every 15 minutes in order to provide data between perturbations; we do this both for low-level (900-500 hPa) and high-level (850-250 hPa) shear. Were the perturbations transient, we would expect to see a sawtooth pattern in the perturbed wind shear in Figure~\ref{fig:f15}, where shear gets nudged lower each hour but then quickly returns back to the shear observed in the forward integration. Instead, low-level shear is consistently lower in the optimization output, confirming that the perturbations to shear are not transient in nature.

\subsection{Water vapor perturbations}
\label{sec:qv}

\begin{figure}[h!]
\includegraphics[width = \textwidth]{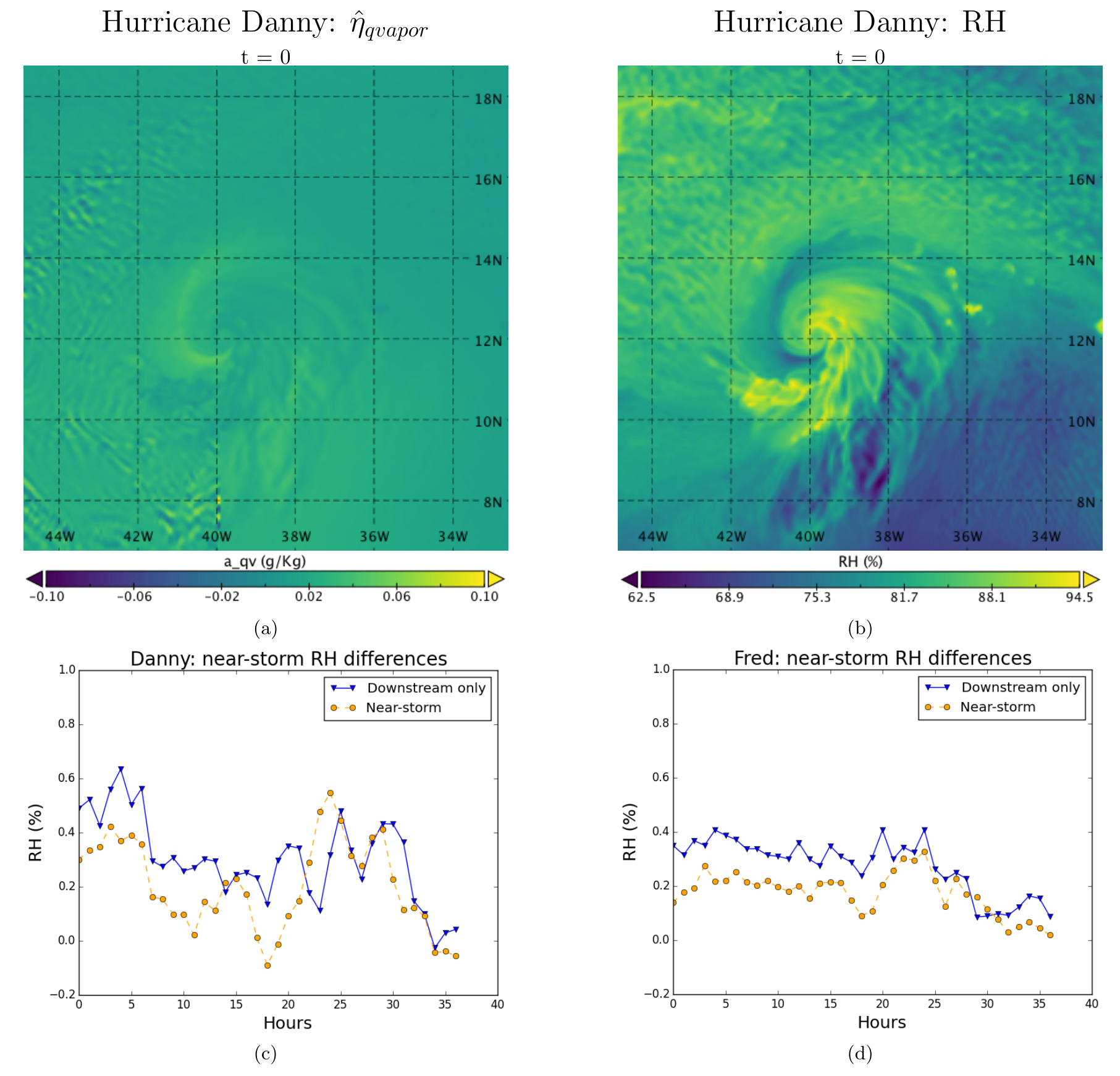}
\caption{Initial time water vapor perturbations (panel a) and initial time relative humidity (panel b) at the surface for Hurricane Danny, as well as time series of near-storm relative humidity differences (optimized minus unperturbed) for the full near-storm domain and only the downstream portion of the domain for Hurricane Danny (panel c) and Hurricane Fred (panel d).}
\label{fig:f17}
\end{figure}

Panels a and b of Figure~\ref{fig:f17} show initial time water vapor perturbations and initial time relative humidity (RH) at the surface for Hurricane Danny, respectively. As expected, the perturbations are generally positive and seem to be larger where relative humidity is smaller. This is particularly visible in the dry area directly south of the storm. In fact, the Pearson product-moment correlation coefficient between $\hat{\boldsymbol{\eta}}_{qvapor}$ and RH is $r = -.61$, confirming that the two are strongly anticorrelated \citep{Ro1988}. Combined with the fact that $\hat{\boldsymbol{\eta}}_{qvapor}$ falls off by nearly two orders of magnitude at distances greater than 500 km from the storm center, this suggests that the major function of these perturbations is to moisten the near-storm environment. This interpretation is broadly consistent with \citet{Di2017}, albeit that study focuses on midlevel relative humidity as opposed to near-surface relative humidity. The water vapor perturbations are qualitatively similar for Hurricane Fred both in that they are confined to the boundary layer and in that they are strongly anticorrelated with relative humidity ($r = -.54$).

Panel a of Figure~\ref{fig:f17} suggests that water vapor perturbations are slightly larger downstream (to the west) of the storm center. One may expect this to be the case, as such perturbations preferentially moisten the air parcels that will shortly spiral inward toward the center of the storm and then convectively rise, thereby leading to more latent heating. Although covariance matrix $B$ contains only instantaneous correlations, the derivative $F'(\cdot)$ can contain dynamic information including the fact that water vapor at an early time in the simulation may become latent heating later on.

To check whether or not action minimization preferentially adds water vapor downstream of the storm, we calculate the average relative humidity in a $500 \times 500$ km box around the storm center in both the action minimization output and the unperturbed integration \citep{Ch2015} and take the difference of the two as a function of time. We then repeat this, but using only the downstream half of the domain. Panels c and d of Figure~\ref{fig:f17} show these differences in near-storm relative humidity for Hurricanes Danny and Fred, respectively. In both cases, the downstream--only average relative humidity shows a larger increase compared to the unperturbed run at most times, suggesting that moistening this portion of the domain is relatively more important for intensification. This result may depend on how exactly a given TC interacts with drier environmental air \citep{Ri2011}.

It is important to note that, compared with temperature and wind perturbations, the action-minimizing water vapor perturbations $\hat{\boldsymbol{\eta}}_{qvapor}$ are fairly small. They comprise approximately $5\%$ and $3\%$ of total running cost for Hurricanes Danny and Fred, respectively (by comparison, temperature perturbations comprise $63\%$ and $45\%$ and wind perturbations comprise $17\%$ and $28\%$ of running cost from Danny and Fred, respectively). In absolute terms, the largest perturbations for Hurricanes Danny and Fred are $0.268$ K and $0.083$ K, respectively, for temperature;  they are $0.188 \mathrm{\ m\ s^{-1}}$ and $0.403 \mathrm{\ m\ s^{-1}}$, respectively, for wind; and they are $0.093 \mathrm{\ g\ kg^{-1}}$ and $0.076 \mathrm{\ g\ kg^{-1}}$, respectively, for water vapor. The relative smallness of the water vapor perturbations is likely because the perturbations are limited almost entirely to the boundary layer (under the 900 hPa level). Only very small amounts of water vapor could be added higher in the atmosphere before reaching saturation (beyond which point the additional water vapor would immediately rain out without affecting storm development); however, such perturbations would incur non-negligible cost since the covariance matrix $B$ accounts for water vapor content decreasing with height. We hypothesize that the algorithm avoids this cost in favor of confining water vapor perturbations to the boundary layer.

To test the importance of the water vapor perturbations, we integrate the model but set $\hat{\boldsymbol{\eta}}_{qvapor} = 0$ at all times. This results in final minimum pressures of $962.7$ hPa for Hurricane Danny and $943.7$ hPa for Hurricane Fred (compared to $962.0$ hPa and $943.3$ hPa, respectively, when using the full perturbations). In both cases, omitting the water vapor perturbations reduces intensification by well under $10\%$ compared to the full perturbation case. The differences in minimum central pressure between runs with and without water vapor perturbations are small enough to be indistinguishable from noise. This shows that, at least for our choice of covariance matrix $B$, perturbations to water vapor are relatively less efficient than perturbations to the temperature and wind fields. It is possible that action minimization chooses to perturb temperature because this allows an indirect increase in moisture via the the Clausius-Clapeyron relationship; thus, perturbations to temperature provide a double boost to intensification.

\subsection{Parameter sensitivity}
\label{sec:sens}

The most important parameter choices in this study are: 1) the frequency of perturbations, 2) the relative weights, $R_i$ and $R_f$, of the running cost and the final cost, and 3) the time horizon. We choose to apply perturbations once an hour over the course of a 36 hour time horizon; however, the results are robust to changes in both of these choices. We show this by varying both time horizon and perturbation frequency at $30$ km resolution and showing that, when the parameters are varied within a reasonable range, both the magnitude and mechanism of intensification remain very similar.

\begin{figure}[h!]
\includegraphics[width = \textwidth]{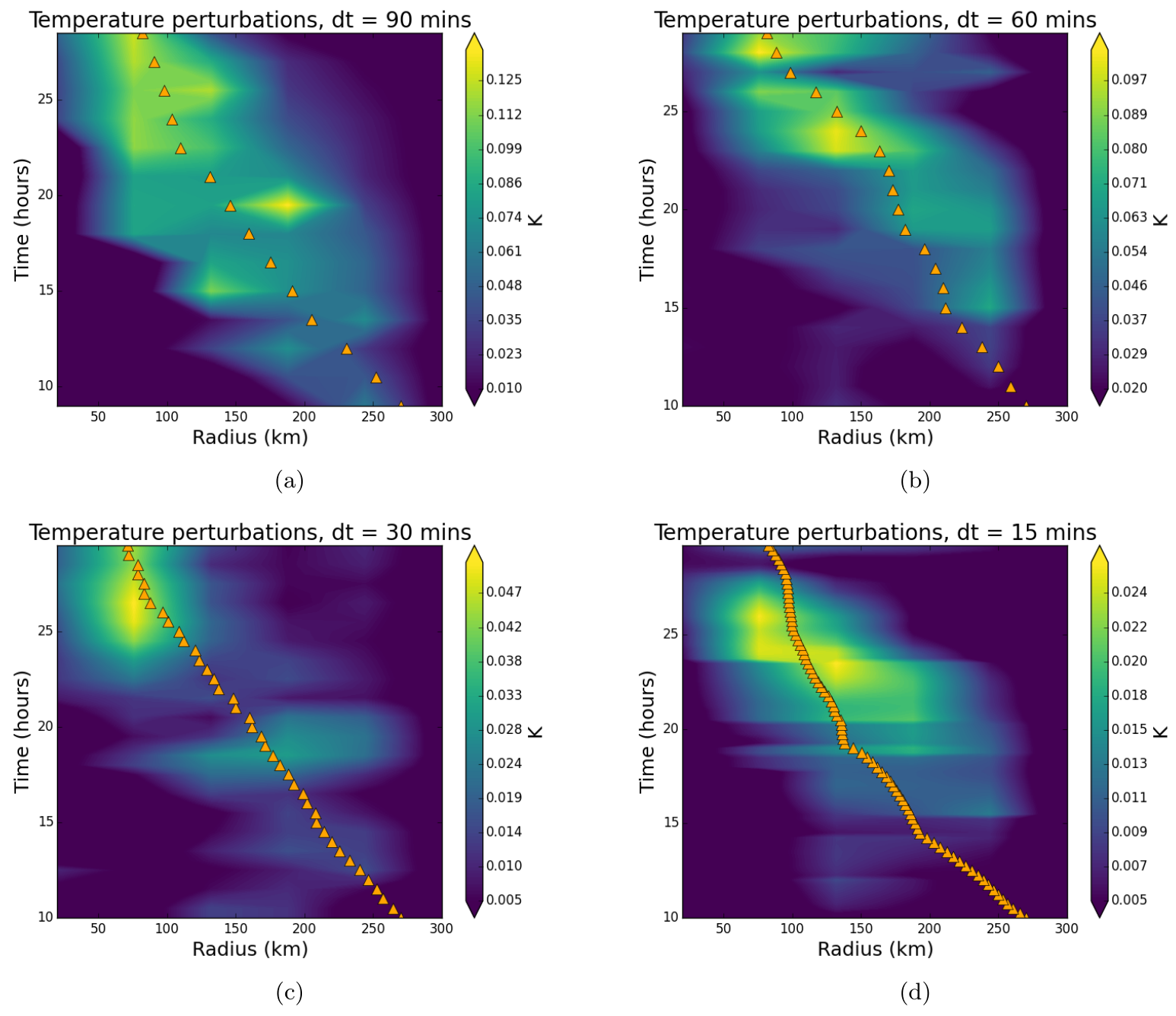}
\caption{Hovm\"{o}ller diagrams of 900 hPa temperature perturbations with radius and time for Hurricanes Danny at various perturbation frequencies. Orange triangles indicate the position of an air parcel advected by mean radial flow. Note that the colorbar scale depends approximately linearly on perturbation frequency.}
\label{fig:f18}
\end{figure}

Figure~\ref{fig:f18} shows Hovm\"{o}ller diagrams of 900 hPa temperature perturbations with radius and time for 30 km resolution simulations of Hurricane Danny; the perturbation frequency varies from 90 minutes to 15 minutes. While the diagrams are not identical, they are qualitatively similar; namely, all show the movement of warm temperature perturbations into the center of the storm with roughly the same velocity as an advected air parcel. The magnitude of the heating at each time step decreases with perturbation frequency so that the total amount of heating remains approximately constant. Further, all of these simulations result in intensification of the pressure minimum by $9.1 - 9.4$ hPa. This suggests that both the magnitude and mechanism of intensification is robust to changes in perturbation frequency.

\begin{figure}[h!]
\includegraphics[width = \textwidth]{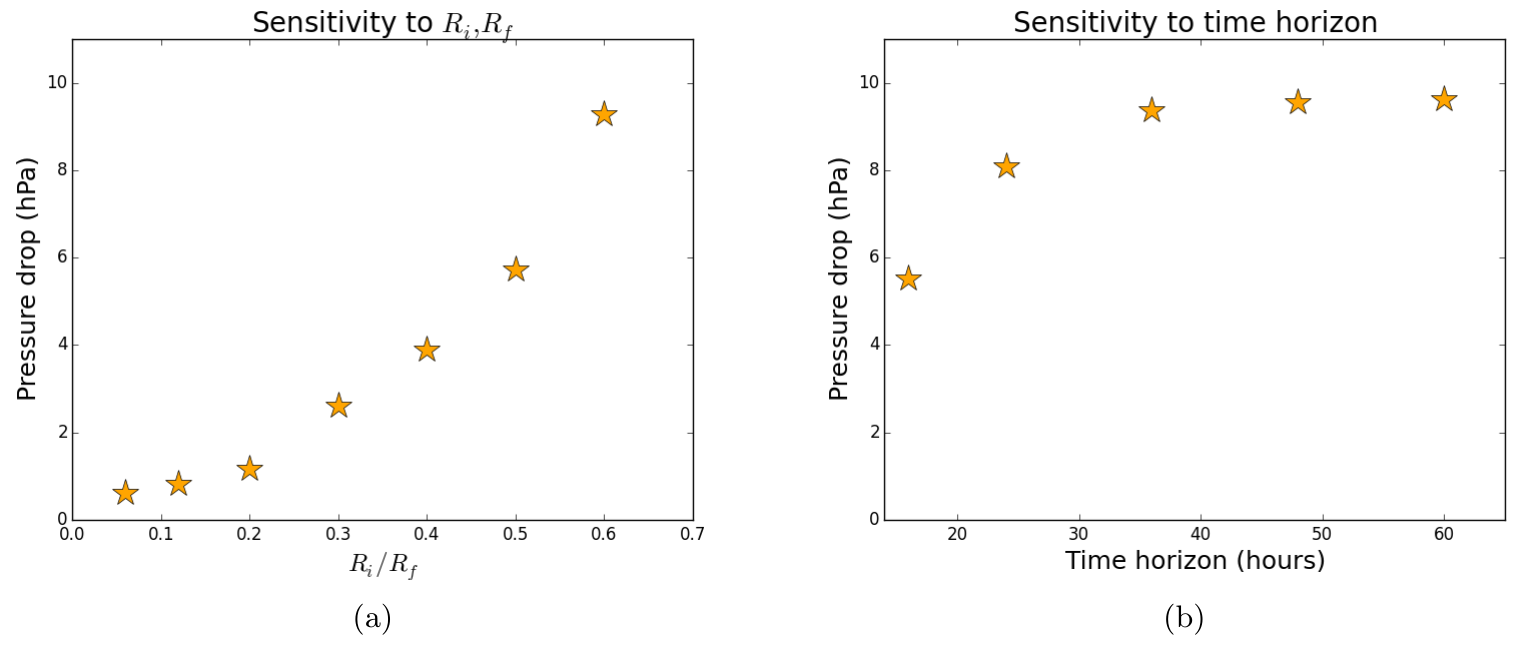}
\caption{Sensitivity of intensification to the ratio of weights $R_i/R_f$ (panel a), and sensitivity of intensification to time horizon (panel b).}
\label{fig:f19}
\end{figure}

We also test the sensitivity of the algorithm to the weights $R_i$ and $R_f$. Since running cost goes as $1/R_f^2$, higher choices of $R_f$ mean relatively less weight is placed on the running cost (and therefore more on the final cost), meaning that we expect higher values of $R_i/R_f$ to result in more intensification. Panel a of Figure~\ref{fig:f19} shows the pressure drop of the action minimization output compared to unperturbed forward integration as a function of the ratio of weights $R_i/R_f$. As expected, the pressure drop increases roughly as the square of $R_i/R_f$. In practice, the high resolution results were obtained by choosing $R_i/R_f$ to be as large as possible while maintaining model stability.

Finally, we run the model for several different time horizons (with the same end time but varying start times). Panel b of Figure~\ref{fig:f19} shows the pressure drop of the action minimization output compared to unperturbed forward integration as a function of time horizon. The magnitude of the pressure drop is nearly constant for time horizons of 36 hours or longer; at 24 hours and below, intensification decreases drastically with time horizon. Further, temperature perturbations for varying time horizon runs (not shown) have the same qualitative behavior as those in Figure~\ref{fig:f18}. This suggests that there is a critical time horizon of approximately 36 hours below which the intensification mechanisms found by action minimization do not have time to maximize intensification; above this threshold, the intensification mechanism and magnitude are relatively unchanged as a function of time horizon.

\section{Discussion}

In this study, we applied an action minimization algorithm to nudge the WRF model into forming more intense TCs than it otherwise would have. For Hurricanes Danny and Fred, our method results in reductions of the minimum pressure of nearly 10 hPa and corresponding increases in maximum surface wind speed of $6-7 \mathrm{\ m\ s}^{-1}$. Action minimization a) causes both storms go from category three to category four on the Saffir-Simpson scale, and b) ensures that both storms undergo RI. In both cases, the nudged model trajectories approach but do not exceed the maximum potential intensities predicted by thermodynamic constraints. This shows that action minimization can consistently result in more intense TCs in weather model output in varying ambient conditions and that these TCs are thermodynamically feasible.

Because extreme events in the tail of the TC intensity distribution cause a disproportionate amount of total TC damage, the study of these events is of particular interest. However, tail events are definitionally difficult to study with traditional ensemble methods since very large ensembles are required to sample these events. We have shown that, typically, action minimization reduces the computational cost of producing extremely intense TCs by approximately an order of magnitude relative to standard ensemble simulation. This makes action minimization a viable tool for studying events in the tail of the TC intensity distribution. Further, the savings in computational cost achieved through action minimization can be used to increase resolution, which may yield insights into rapid intensification in future studies.

For both Hurricanes Danny and Fred, the action-minimizing perturbations to the forward model introduce low-level warm temperature anomalies that move into the warm cores of the storms in a manner consistent with advection. These anomalies have maxima at approximately $z = 2$ km, and suggest that low-level convection may be a preferred mechanism of TC intensification in the WRF model. This provides evidence for recent observational studies that argue that shallow to moderate convection can be a trigger for RI, while deep convection is often a response to RI \citep{Ch2013, Ta2015}.

The temperature perturbations for Hurricane Danny are asymmetrical in such a way as to make the temperature field more asymmetrical during the middle of the optimization time horizon. As expected, the vortex axisymmetrizes by the end of the optimization time horizon. However, asymmetry of perturbations contributes to intensification since integrating the model with only azimuthally averaged perturbations results in an approximately $30\%$ reduction in intensification as compared to integrating using the full perturbations. Further, and in agreement with \citet{No2007}, integrating with only the purely asymmetric component of the perturbations results in a slight weakening of the vortex. Our results differ markedly from that study because, at least for the specific set of perturbations produced by action minimization, both asymmetry and non-linearity of the response of the vortex to warming significantly increase intensification.

Our results for Hurricane Danny may differ from those of previous studies due to the specific nature of perturbations applied to the vortex. The significance of nonlinearity may arise from the fact that the asymmetric velocities in this study are above the ``limit of where we can expect linear calculations to be accurate'' described in \citet{No2007}. Further, the action-minimizing perturbations differ from those used in previous studies in that they a) are centered at lower altitudes, b) have time-dependent radial structure, and c) have azimuthal structure that is not constrained by a particular functional form. This is a major strength of the action minimization algorithm, as it is able to search over the space of all possible perturbations whereas previous studies and their conclusions were limited to the specific perturbations that they imposed. Finally, the fact that asymmetry of perturbations does not significantly impact the intensification of Hurricane Fred suggests that the role of asymmetric heating depends on pre-existing asymmetry and storm structure; this result is consistent with \citet{Mo2005}. 

For both Hurricanes Danny and Fred, the action-minimizing perturbations also reduce low-level environmental wind shear. The decrease in low-level shear is approximately $\sim$1 m\ s$^{-1}$ for Fred at the beginning of the integration and is about five times larger than the reduction in shear for Hurricane Danny. This may be because the unperturbed Danny integration already starts below a postulated threshold near 4 m\ s$^{-1}$ below which reduction in shear has little effect. While previous studies \citep[e.g.][]{Ch2013} suggest that environments where shear is too low ($<5$ m\ s$^{-1}$) are actually detrimental to TC intensification as compared to moderate shear ($5-10$ m\ s$^{-1}$) environments, our results are more in line with those of \citet{No2012}, who show that extremely low-shear environments are no less favorable to intensification than are moderate-shear environments but that the effect of reducing shear gets smaller as the magnitude of the shear gets smaller.

Finally, action-minimizing perturbations to water vapor moisten the near-storm environment. These perturbations are strongly anticorrelated with relative humidity, and are almost entirely confined to the boundary layer. Further, relative humidity is preferentially increased downstream of the storm in the action minimization output. Somewhat surprisingly, the perturbations to water vapor are relatively small compared to perturbations to the wind and temperature fields, both in terms of the running cost allocated to them and in terms of their effect on the final minimum pressure. 

While these results show that the trajectories resulting from action minimization can elucidate the physical trajectories by which TCs undergo rapid intensification in the WRF model, our work has several important limitations. First, action minimization requires a specific model and can only contribute to understanding of trajectories/processes within that model. If a model is incapable of simulating the true process by which an event occurs, action minimization will not yield that true trajectory; instead, action minimization should be thought of as a method for efficiently accessing and studying tail events that a model can produce.

Second, the choice of covariance matrix $B$ may constrain the output that action minimization can produce. Specifically, we generate $B$ from long model runs that mostly access quiescent states. The flows in these states are in balance and relatively smooth, so it is possible that our choice of $B$ imposes smoothness and overly penalizes large gradients. Such gradients are uncommon in most weather states but prevalent under TC conditions. While the choice of $B$ poses some difficulty, we also note that any ensemble algorithm also requires an analogous choice in how to generate ensemble members that diverge from one another. This is true even for ensembles where only the initial condition is varied, as choosing the different initial conditions requires drawing them from a distribution that is characterized by some covariance matrix $B$.

Third, and as with ensemble methods, diagnosing the causes of rapid TC intensification requires an analysis of trajectories.  If one has a putative mechanism in mind, one can assess the degree to which the action-minimized path agrees with that mechanism, as in our analysis of the effects of asymmetries in the perturbations.  Alternatively, one can systematically investigate a potential mechanism by modifying the background covariance matrix to penalize certain classes of perturbations (e.g. asymmetric perturbations) and compare the resulting trajectory to the trajectory found with the original background covariance.  Identification of new mechanisms from the trajectories resulting from action minimization (or ensemble simulation) is a more difficult problem.  It can, in principle, be approached using standard data analysis tools like PCA \citep{Br1992} or diffusion maps \citep{Co2008, Gi2012, Pl2014}.  However, the data are both high dimensional and time dependent and effective analysis would benefit from the development of new tools capable of distinguishing important features of the trajectories.

Fourth, the optimal perturbations could depend on both resolution and on choice of model. While the results presented in section~\ref{sec:sens} show little qualitative dependence on resolution, the action-minimizing perturbations could change at a resolution that allows the model to resolve convection. In both this study and in previous work, temperature perturbations have often been considered as a proxy for latent heating that results from convection. We are not aware of studies that impose convective perturbations in a more direct way, so it is difficult to predict how doing so would affect the results. Further, the choice of model and of parametrizations (convective and boundary layers schemes, etc.) are significantly limited in this study, since WRF is one of few publicly available climate or weather models with adjoint capabilities. Within the WRFPLUS adjoint scheme, only several state variables can be perturbed and only a small subset of all parametrizations are available. It is possible that applying action minimization to a different model with different parametrizations, for example to a coupled GCM, would yield different results. This study is necessarily constrained by the imperfections of the WRF model and by the approximations inherent in the adjoint scheme.

The introduction of action minimization into the repository of tools for the study of rapid intensification calls for several directions of future work. First, we intend to apply action minimization to a larger sample of historical TCs to better study intensification mechanisms. By economizing computation through the use of action minimization, these simulations can be obtained at higher resolution than previously feasible. Such a sample will show what it would have taken to cause historical storms to further intensify and will allow us to quantify whether or not certain pathways of intensification are consistently likelier than others. 

By asking what it takes to cause further intensification, both this study and the ones proposed above seek to elucidate the processes that are sufficient for rapid intensification. Action minimization can also be used to investigate necessary processes and conditions. To do so, we can modify the action functional used in this study to instead prefer final states with quiescent weather (instead of intense TCs). By doing so, we ask the converse of the question that we study here, namely: what would it take to prevent intensification? By combining the study of both questions, we can provide insight into both the necessary and sufficient processes required for rapid intensification in the WRF model.

\section{Conclusions}

In this study, we apply an action minimization algorithm to Hurricanes Danny and Fred using the WRF and WRFPLUS models to reach the following conclusions:
\begin{enumerate}
\item{Action minimization can consistently cause TCs in the WRF model to intensify beyond their maximum intensities attained via forward integration.}
\item{Compared to ensemble simulation, action minimization yields computational savings of approximately an order of magnitude in accessing extreme events at the tail of the TC intensification distribution.}
\item{Model trajectories output by action minimization are consistent with thermodynamic constraints, as evidenced by the fact that these trajectories have TCs intensifying close to but not beyond their maximum potential intensities.}
\item{Intensification pathways include low-level heating that moves toward the center of the storm. Depending on pre-existing storm structure, the asymmetry of the heating can play a non-negligible role in intensification. Further, intensification responds nonlinearly to the symmetric and asymmetric components of this heating.} 
\item{Action minimization preferentially reduces low-level vertical shear compared to high-level vertical shear, especially when low-level shear is above a threshold of $\sim5$ m\ s$^{-1}$. This leads to intensification by reducing storm tilt.}
\item{Action minimization has several applications in the study of TCs, including a) the study of rapid intensification mechanisms and b) reduction of the computational cost required to generate high resolution simulations to study both necessary and sufficient conditions for rapid intensification.}
\end{enumerate}

In sum, we have demonstrated the feasibility of action minimization as a tool for the study of TC rapid intensification. Further, we have provided evidence for the roles of low-level wind shear and of low-level, asymmetric heating in rapid intensification. By producing high resolution trajectories in the tails of the TC intensification distribution for many storms, action minimization can continue to aid investigation of rapid intensification mechanisms in future studies.

\acknowledgments

We acknowledge support from the National Science Foundation under NSF award number 1623064. This work was supported by the Department of Energy Computational Science Graduate Fellowship Program of the Office of Science and National Nuclear Security Administration in the Department of Energy under contract DE-FG02-97ER25308. RW and JW are supported by the Advanced Scientific Computing Research Program within the DOE Office of Science through award DE-SC0014205. RW was also supported by NSF RTG award number 1547396. MO was supported by the T. C. Chamberlin Postdoctoral Fellowship at the University of Chicago. This work was completed with resources provided by the University of Chicago Research Computing Center. We would like to thank four anonymous reviewers, as well as Dr. David Nolan, for extensive suggestions for improving the manuscript. Code developed for action minimization applied to the WRF model is available at https://knowledge.uchicago.edu/handle/11417/1091.

\listofchanges

\end{document}